\documentclass[5p,twocolumn,times]{elsarticle}
\usepackage{tikz}
\usetikzlibrary{arrows}
\usetikzlibrary{positioning}
\usepackage{amssymb}
\usepackage{amsmath}
\usepackage{color}
\usepackage{xspace}
\usepackage{subfigure}
\usepackage{wrapfig}
\usepackage{algorithm}
\usepackage{algorithmic}
\usepackage{multirow}
\usepackage{colortbl}
\usepackage{textcomp} % for the old style numbers
\usepackage{listings}
\lstdefinelanguage{CUDA}[]{C++}{
    morekeywords={__global__,__device__,__shared__,__syncthreads,threadIdx,blockIdx,float3,float4,rsqrtf},
}
\definecolor{listinggray}{gray}{0.5}
\definecolor{keycolor}{rgb}{0.459,0.071,0.294}
\definecolor{linkblue}{rgb}{0,0.26,0.46}
\definecolor{cblue}{rgb}{0.098,0.357,0.675}
\usepackage[
    pdftex,
    pdftitle={How to obtain efficient GPU kernels: an illustration using FMM \& FGT},
    pdfauthor={F.~A. Cruz, S.~K. Layton, L.~A. Barba},
    pdfpagemode={UseOutlines},
    bookmarks, bookmarksopen,bookmarksnumbered={True},
    colorlinks, linkcolor={linkblue},citecolor={linkblue},urlcolor={linkblue}
    ]{hyperref}

\makeatletter
\def\url@leostyle{%
  \@ifundefined{selectfont}{\def\UrlFont{\sf}}{\def\UrlFont{\footnotesize\sffamily}}}
\makeatother
\newcommand{\opencl}{\textsc{O}pen\textsc{CL}\xspace}

\newcommand{\uKern}{\mathbb{K}}
\newcommand{\bigON}{$\mathcal{O}(N)$}
\newcommand{\bigOsq}{$\mathcal{O}(N^2)$}
\newcommand{\fmm}{\textsc{fmm}}
\newcommand{\fgt}{\textsc{fgt}}

\newcommand{\rbf}{\textsc{rbf}}
\newcommand{\cfd}{\textsc{cfd}}

\newcommand{\cpu}{\textsc{cpu}}
\newcommand{\gpu}{\textsc{gpu}}
\newcommand{\cuda}{\textsc{cuda}}
\newcommand{\fermi}{\textsc{fermi}}
\newcommand{\tesla}{\textsc{tesla}}
\newcommand{\tpc}{\textsc{tpc}}

\newcommand{\ME}{\textsc{me}}
\newcommand{\LE}{\textsc{le}}
\newcommand{\ML}{\textsc{m}\texttwooldstyle\textsc{l}\xspace} % M2L
\newcommand{\NV}{\textsc{nvidia}}
\newcommand{\sm}{\textsc{sm}}

\newlength{\hup}
\begin{document}

\begin{frontmatter}

% Title, authors and addresses

% use the thanksref command within \title, \author or \address for footnotes;
% use the corauthref command within \author for corresponding author footnotes;
% use the ead command for the email address,
% and the form \ead[url] for the home page:
% \title{Title\thanksref{label1}}
% \thanks[label1]{}
% \author{Name\corauthref{cor1}\thanksref{label2}}
% \ead{email address}
% \ead[url]{home page}
% \thanks[label2]{}
% \corauth[cor1]{}
% \address{Address\thanksref{label3}}
% \thanks[label3]{}

\title{How to obtain efficient GPU kernels: an illustration using FMM \& FGT algorithms}

% use optional labels to link authors explicitly to addresses:
% \author[label1,label2]{}
% \address[label1]{}
% \address[label2]{}

\author[fac]{Felipe A. Cruz}
\ead{f.cruz@bristol.ac.uk}

\author[lab]{Simon K. Layton}
\ead{slayton@bu.edu}

\author[lab]{L. A. Barba\corref{bu}} 
\ead{labarba@bu.edu}
\address[fac]{Department of Mathematics, University of Bristol}
\address[lab]{Mechanical Engineering Department, Boston University}
\cortext[bu]{Corresponding author.  Address: 110 Cummington St, Boston MA 02215}

\begin{abstract}

Computing on graphics processors is maybe one of the most important developments in computational science to happen in decades.  Not since the arrival of the Beowulf cluster, which combined open source software with commodity hardware to truly democratize high-performance computing, has the community been so electrified.  Like then, the opportunity comes with challenges.  The formulation of scientific algorithms to take advantage of the performance offered by the new architecture requires rethinking core methods. Here, we have tackled fast summation algorithms (fast multipole method and fast Gauss transform), and applied algorithmic redesign for attaining performance on {\gpu}s.  The progression of performance improvements attained illustrates the exercise of formulating algorithms for the massively parallel architecture of the {\gpu}.  The end result has been {\gpu} kernels that run at over 500 Gop/s on one {\NV} {\tesla} C1060 card, thereby reaching close to practical peak. 

\end{abstract}

\begin{keyword}
% keywords here, in the form: keyword \sep keyword
fast summation methods \sep fast multipole method \sep fast Gauss transform \sep heterogeneous computing
% PACS codes here, in the form: \PACS code \sep code
%\PACS 
\end{keyword}
\end{frontmatter}

%% TEXT BODY

%% SECTION 1 -- INTRODUCTION
\section{Introduction}
\label{intro}

%!TEX root = CruzLaytonBarba2010.tex
%%   intro.tex

%%% SECTION

The reality of the past few years is that the computing industry is betting everything on parallel computing, with first the multi-core era, and now a many-core trend \cite[]{asanovic2006}.  The implication of this trend is that it will not be enough to develop applications for quad- or eight-core systems, but rather for tens and hundreds of processor systems.  The unwelcome news that industry players are delivering is that many-core architectures require ``going back to the algorithmic drawing board''\cite[]{Ghuloum2008}.  Incrementally thinking about ``parallelizing'' an existing serial formulation of an algorithm will not work for the many-core architecture. We have to \emph{rethink} the logic of the formulation from the bottom up, and recast the algorithm starting from the mathematics.

\begin{figure}
\centering\vspace{4mm}
    \includegraphics[width=0.40\textwidth]{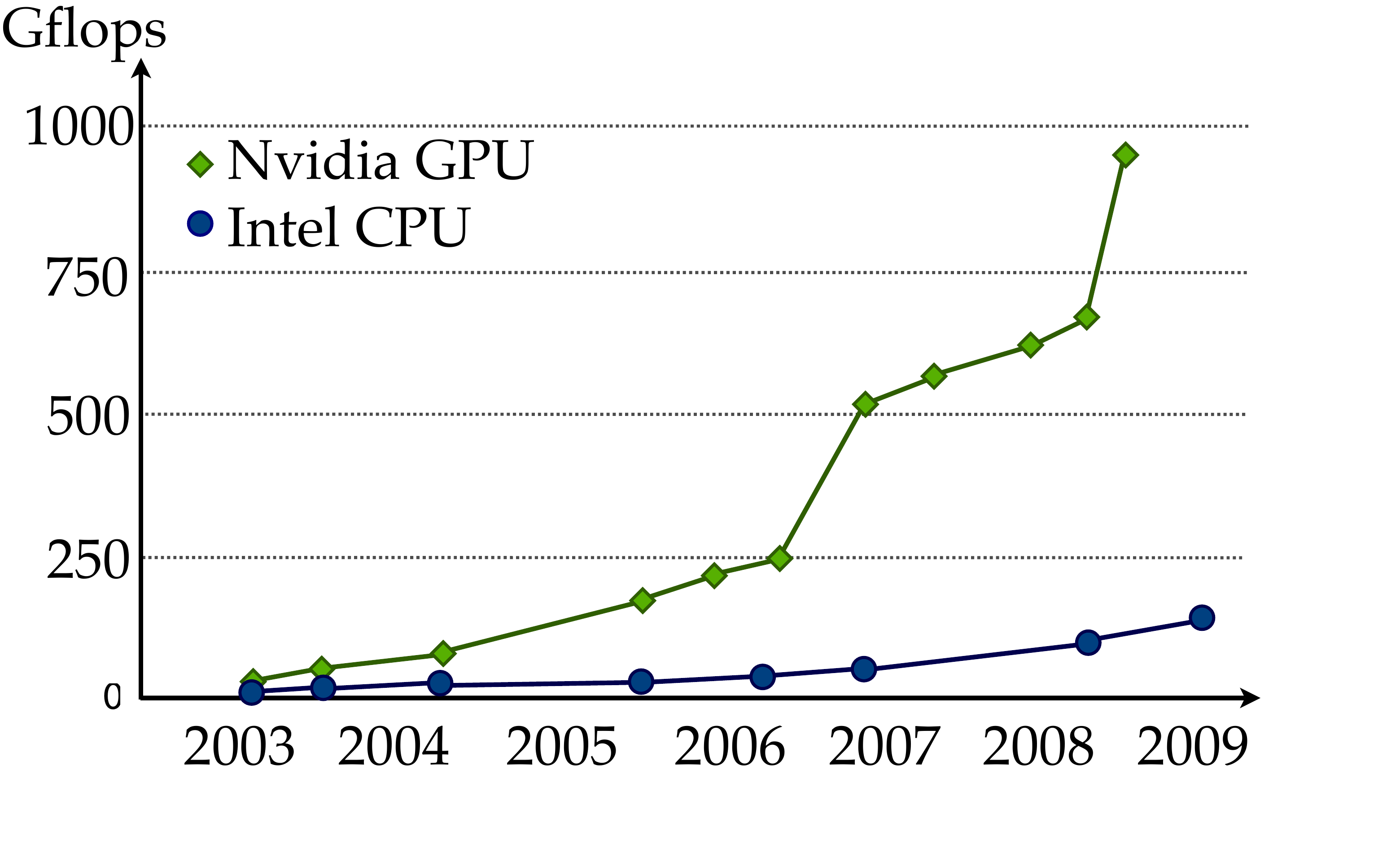}
\caption{Evolution of the computing capacity of several generations of {\cpu}s and {\gpu}s.  Adapted from \cite{cuda-guide}.}\vspace{-2mm}
\label{fig:cpuspeed}
\end{figure}

Concurrently to the changing situation for {\cpu} technology in recent years, new prominent hardware architectures have come into play.  Attracting a tremendous amount of the attention is the programmable {\gpu} (graphics processing unit), with its dramatically faster progress compared to {\cpu}s. 
Figure \ref{fig:cpuspeed} shows the computing capacity, measured in Gigaflop/s, of subsequent generations of Intel {\cpu}s and \NV\ {\gpu}s, as shown in \cite{cuda-guide} and oft-times cited since.  The peak performance of the latest {\gpu} is several times greater than the latest {\cpu}, but more important is the trend.  In addition to the raw performance, {\gpu}s have excellent performance-per-watt.  As high-performance computing centers are more and more burdened by high power costs, the energy efficiency of the {\gpu} makes it an irresistible choice.    The price/performance ratio, finally, makes the {\gpu} architecture a winner in various settings.  Given that the market for {\gpu}s was driven by the games industry, it is commodity hardware and thus cheaper than alternatives for high-performance computing based on `mainframe' servers.

The real opportunity for using {\gpu}s in scientific computing came about with the release of {\cuda}, a C-language extension that gave programmers relatively easy access to the {\gpu} for coding general algorithms %
(first publicly released by \NV\ in February 2007).   
However, there are significant challenges as the specialized hardware architecture again forces us to go ``back to the algorithmic drawing board'' \cite{Ghuloum2008}.
Indeed, there is an immediate need for research into algorithms so that we can exploit the performance offerings of the {\gpu} architecture.  This is the motivation of the present work.

In scientific computing, there are multiple algorithms that pervade through a variety of scientific or engineering applications.  
%These fundamental algorithms need to be investigated to discern how they map to the massively parallel architecture of {\gpu}s.  There are preliminary indications that some algorithms will be better suited to exploit the {\gpu} architecture than others.  
Here, we focus on two fast summation algorithms\,---fast multipole method, \fmm, and fast Gauss transform, \fgt--- and present examples of the workflow to produce a formulation of these algorithms that will maximize the performance on the \gpu\ hardware architecture.
%% Here make a reference to the 7 dwarfs, and cite them. Open for the importance of fast summation methods... this is the beginning of the transition to the relevant problem.
The algorithms selected in this work program constitute enabling mathematical technology used in many numerical methods.  These, in turn, are key tools for enabling specific application areas in science and engineering.  

The focus of our research in regards to the applications is in large-scale computational fluid dynamics (\cfd) using particle methods, and radial basis function ({\rbf}) methods.  
For instance, consider the vortex particle method in \cfd, where one has a large number of particle-like objects that are used for the mesh-free discretization of the Navier-Stokes equation in vorticity formulation; see \cite{barbaETal2005} for details of the method. The evaluation of the velocity field in this formulation results in an $N$-body problem.  In general, the $N$-body problem consists of evaluating all the pair-wise interactions of a set of $N$ bodies due to a force potential, given by:
\begin{equation}
f(x_{j}) = \sum_{i = 1}^{N} c_{i} \, \uKern(x_{j} , x_{i}).
\label{eq:n_body}
\end{equation}

\noindent Here, $\{(c_{i}, x_{i})\}$ represents the set of $N$ source particles with each particle $i$ having a coordinate $x_{i}$ and a weight $c_{i}$; and  $\uKern(x_{j} , x_{i})$ represents the kernel function that defines the interactions between two particles. The direct evaluation of Equation~\ref{eq:n_body} at all $N$ particle locations has a computational complexity of \bigOsq. It is often the case in problems of scientific interest that the set of particles used in simulations is in the order of thousands to millions. For such large simulations, the direct evaluation of the particle interactions would take an unreasonable amount of computational time. This is one of the main motivations for the use of fast summation methods that are able to accelerate the evaluation of the particle interactions. Such methods reduce the computational complexity of the $N$-body problem to $\mathcal{O}(N \log N)$ or \bigON, by rapidly computing and aggregating approximated pairwise particle interactions.

Perhaps the most renowned of the fast $N$-body methods is the fast multipole method, \fmm, discovered by Greengard and Rokhlin \cite{GreengardRokhlin1987}. Its applications have included long-range electrostatics in \textsc{dna} simulations \cite{FenleyETal1996}, calculations of vortex sheet evolution \cite{HamiltonMajda1995}, room acoustics and scattering from multiple bodies by means of the Helmholtz equation \cite{GumerovDuraiswami2004}, and many others.  We will describe the basic aspects of the \fmm\ algorithm in the next section, but in a necessarily brief fashion. Another fast algorithm, which specializes the \fmm\ to the case when the kernel function in \eqref{eq:n_body} is a Gaussian function, is the fast Gauss transform, \fgt\ \cite{GreengardStrain1991}.  This paper covers both the \fmm\ and the \fgt, and their implementation on the \gpu\ architecture for achieving maximum performance.  After giving a background on the two algorithms that we treat, a motivation for \gpu\ computing follows in \S\ref{motivation}.  The optimization of the algorithmic kernels is described in detail in \S\ref{gpuformulation}, followed by details of the implementation and discussion of the performance obtained (\S\ref{implementation}) and final concluding remarks.

%% SECTION 2
\section{Background on fast summation methods} \label{background}

%!TEX root = CruzLaytonBarba2010.tex
%%   fmm_background.tex

%%% SECTION
%\section{Background on fast summation methods}
%\label{background}

%%%  SUBSECTION
Of central importance to fast summation methods is the idea that when evaluating pairwise interactions, the solution is only required up to a given accuracy. This idea opens the door to methods that use efficient schemes to rapidly compute an approximation to the interactions, as an alternative to directly evaluating them. It can be said that fast summation methods trade numerical accuracy for computational speed. In this work, we focus on two methods that are classified as analytic-based fast algorithms, as they use analytic approximations to obtain speed and arbitrary accuracy: the Fast Multipole Method, \fmm, and the closely-linked Fast Gauss Transform, \fgt.

\subsection{Fast multipole method}\label{subsec:fmm_back}

The {\fmm} was developed to calculate the pairwise interactions of large sets of particles. Whenever the physical interactions of interest have long-range effects (\emph{e.g.}, gravitational and electrostatic problems), evaluating the interaction among $N$ particles in principle requires $\mathcal{O}(N^2)$ operations.  The {\fmm} of Greengard and Rokhlin~\cite{GreengardRokhlin1987} reduces the computational complexity of this problem to $\mathcal{O}(N)$ operations by approximating and aggregating the pairwise interactions. A detailed description of the {\fmm} algorithm is outside the scope of this paper. We will only give a brief description of the method and outline the most important steps of the algorithm. For a formal and more detailed description of the {\fmm} algorithm, the reader can consult the original reference \cite{GreengardRokhlin1987}, or other surveys such as \cite{BeatsonGreengard97}.

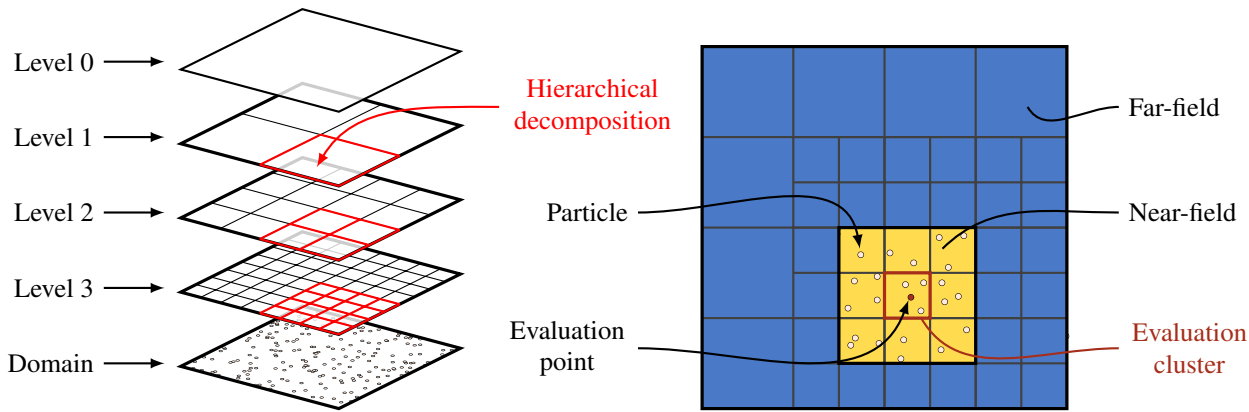
\begin{figure*}
    %% Diagram for the hierarchical decomposition of the space.
% The diagram shows the different levels of refinement and
% the near-field and far-field coverage.

%% Python-code for particle generator:
%from numpy import *
%n = 200
%data = random.uniform(low=0.01,high=3.99,size=(n,2))
%for i in range(n):
%  print "(%.2f,%.2f)," % (data[i][0], data[i][1]),

% Latex MACRO for plotting particles
\newcommand{\DrawParticles}{
        \foreach \position in {
(3.82,0.08), (2.43,3.62), (1.54,2.77), (0.91,3.26), (1.38,1.64), (2.22,3.52), (2.54,2.22), (1.81,2.87), (0.72,0.89), (1.17,1.38), (1.53,0.16), (3.70,2.05), (2.22,0.44), (0.06,3.09), (2.93,3.93), (2.86,0.02), (1.99,2.66), (2.63,1.39), (3.67,3.76), (3.03,3.65), (2.11,3.21), (3.83,3.09), (1.23,0.89), (3.52,0.09), (3.45,3.52), (2.90,3.79), (2.73,0.40), (1.29,1.65), (3.61,2.20), (1.81,0.40), (0.23,1.31), (2.18,0.55), (1.92,1.20), (3.13,3.34), (1.36,1.67), (0.75,0.43), (1.28,3.24), (3.24,2.71), (2.78,0.47), (2.87,1.91), (1.87,3.76), (2.85,3.53), (3.39,3.01), (1.59,2.83), (2.06,1.72), (2.80,3.08), (2.40,1.08), (3.96,0.99), (0.62,1.91), (3.90,1.18), (3.78,3.06), (0.28,0.65), (1.00,2.37), (3.60,0.85), (0.37,0.41), (3.82,1.44), (0.11,2.09), (1.68,0.78), (1.48,1.93), (2.33,0.39), (0.67,3.72), (0.06,1.78), (1.65,2.10), (0.24,2.48), (1.60,1.11), (3.19,1.86), (1.38,1.94), (1.46,0.77), (0.46,0.04), (3.19,2.57), (2.01,3.10), (2.23,1.37), (2.70,1.56), (1.15,0.80), (2.11,2.23), (2.79,0.32), (3.56,3.34), (1.63,0.70), (1.67,1.41), (0.37,2.04), (1.03,2.17), (2.31,3.08), (3.36,3.60), (1.48,1.98), (2.50,2.98), (2.08,2.60), (3.30,3.45), (3.73,2.73), (3.67,2.94), (3.28,1.50), (2.51,2.08), (2.90,0.87), (0.15,2.66), (3.28,0.47), (2.35,2.49), (1.68,2.20), (1.53,0.46), (3.04,3.51), (0.17,3.26), (3.86,1.96), (2.81,1.24), (0.26,2.73), (3.99,0.80), (2.37,0.10), (3.68,3.78), (3.43,2.23), (3.71,2.35), (1.94,0.72), (3.07,3.90), (0.73,2.79), (3.02,0.07), (0.20,0.81), (2.62,0.65), (0.61,2.73), (2.94,3.83), (0.74,3.31), (2.58,0.18), (0.71,0.09), (1.27,3.86), (3.51,3.71), (2.79,3.85), (0.52,2.79), (2.66,1.18), (3.47,1.45), (3.07,3.17), (2.27,3.55), (3.18,1.69), (3.83,1.25), (3.02,1.33), (0.81,3.29), (2.32,1.61), (3.72,0.22), (1.25,1.36), (0.30,0.17), (3.93,0.96), (3.48,1.48), (3.19,3.82), (3.16,0.61), (1.29,1.74), (2.50,0.33), (0.66,3.92), (0.84,0.33), (3.85,1.16), (0.67,1.46), (1.27,2.19), (0.06,2.15), (2.83,0.29), (2.43,1.39), (1.35,2.45), (2.20,2.21), (0.23,1.85), (1.97,0.28), (0.08,3.53), (2.86,2.96), (2.37,2.45), (3.93,3.45), (1.75,3.05), (0.61,1.77), (2.29,3.44), (2.00,3.80), (0.36,3.92), (0.11,3.28), (0.06,3.73), (3.59,1.02), (3.91,1.48), (3.04,2.55), (1.66,3.44), (0.79,2.70), (3.01,2.26), (3.77,2.29), (2.42,3.83), (2.36,0.43), (1.27,1.09), (1.77,2.96), (1.92,1.46), (1.18,0.24), (2.58,3.56), (0.93,2.95), (2.21,0.76), (1.74,1.70), (1.24,3.04), (1.96,2.10), (0.86,1.12), (1.12,3.85), (2.16,0.07), (2.37,2.97), (3.64,1.37), (3.69,0.78), (1.38,0.68), (3.22,3.31), (3.91,3.54), (0.42,2.63), (1.30,3.91), (2.22,3.75), (0.40,1.15), (2.60,1.89), (2.12,2.20)
        } {
            \draw \position circle (0.03cm);
            \fill[particleColor!20] \position circle (0.03cm);}
}
% end of MACRO

 %% Diagram of the hierarchical decomposition and clustering 
\begin{tikzpicture}[scale=0.4]
% help line, erase when finished
%\draw[help lines] (-4,0) grid (38,14);

% Define colors to use
\definecolor{ilistColor}{rgb}{0.29, 0.48, 0.78}
\definecolor{nearFieldColor}{rgb}{1.0,0.86,0.31}
\definecolor{particleColor}{rgb}{1.0,0.59,0.31}
\definecolor{boxColor}{rgb}{0.65,0.18,0.10}

% Figure of the hierarchical domain decomposition image
  \begin{scope}[xshift=200]
    
    % Domain
    \begin{scope}[yshift=0,yslant=0.5,xslant=-1.3]
        \fill[white,fill opacity=0.8] (0,0) rectangle (4,4);
        % Add particles to the plot
        \DrawParticles
         % Borders of the domain
        \draw[black,very thick] (0,0) rectangle (4,4);
    \end{scope}
    
    % Level 3
    \begin{scope}[yshift=70,yslant=0.5,xslant=-1.3]
        \fill[white,fill opacity=0.8] (0,0) rectangle (4,4);
        \draw[step=0.5, black] (0,0) grid (4,4); %defining grids
        \draw[black,very thick] (0,0) rectangle (4,4);%marking borders
        \draw[step=0.5, red, thick] (0,0) grid (2,2); %Active cluster border
    \end{scope}
    
    % Level 2
    \begin{scope}[yshift=140,yslant=0.5,xslant=-1.3]
        \fill[white,fill opacity=0.8] (0,0) rectangle (4,4);
        \draw[step=1, black] (0,0) grid (4,4); %defining grids
        \draw[black,very thick] (0,0) rectangle (4,4);%marking borders
        \draw[step=1, red, thick] (0,0) grid (2,2); %Active cluster border
    \end{scope}
    
    % Level 1
    \begin{scope}[yshift=210,yslant=0.5,xslant=-1.3]
        \fill[white,fill opacity=0.8] (0,0) rectangle (4,4);
        \draw[step=2, black] (0,0) grid (4,4); %defining grids
        \draw[black,very thick] (0,0) rectangle (4,4);%marking borders
        \draw[step=2, red, thick] (0,0) grid (2,2); %Active cluster border
    \end{scope}
    
    % Level 0
    \begin{scope}[yshift=280,yslant=0.5,xslant=-1.3]
        \fill[white,fill opacity=0.8] (0,0) rectangle (4,4);
        \draw[black, thick] (0,0) rectangle (4,4);%marking borders
    \end{scope}
  
    \begin{scope}[xshift=-220]
      % Adding labels
      \draw[-latex,thick] (0,11.5) node[left]{Level $0$}  to (2.0,11.5);
      \draw[-latex,thick] (0,9.0) node[left]{Level $1$}  to (2.0,9.0);
      \draw[-latex,thick] (0,6.5) node[left]{Level $2$}  to (2.0,6.5);
      \draw[-latex,thick] (0,4.0) node[left]{Level $3$} to (2.0,4.0);
      \draw[-latex,thick] (0,1.5) node[left]{Domain}    to (2.0,1.5);
    \end{scope}
  
    \begin{scope}[xshift=150]
      % Adding labels
      \draw[-latex,thick,red] (0,10.0) node[right, text width=2.2cm, text centered]{Hierarchical decomposition}  to [out=180,in=60] (-6.0,8.0);
    \end{scope}
  \end{scope}

%% Near-field and far-field
  \begin{scope}[xshift=540]
  
    \begin{scope}[scale=3.0]
        \fill[nearFieldColor] (1.5,0.5) rectangle (3.0,2.0); % BG-coloring Near-field
        \DrawParticles % Set of particles
        % Evaluation point
        \draw (2.29,1.23) circle (0.03cm);
        \fill[boxColor] (2.29,1.23) circle (0.03cm);
        % Color interaction list after particles
        \fill[ilistColor] (0,2) rectangle (4,4);
        \fill[ilistColor] (0,0) rectangle (1.5,2.0); 
        \fill[ilistColor] (0,0) rectangle (4.0,0.5); 
        \fill[ilistColor] (3,0) rectangle (4,2); 
        % Space subdivision
        \draw[step=1, black!75, thick] (0,0) grid (4,4); % Level 2 clusters
        \draw[step=0.5, black!75, thick] (1,0) grid (4,3); % Level 3 clusters
        \draw[black,very thick] (0,0) rectangle (4,4);% Domain borders
        \draw[black,very thick] (1.5,0.5) rectangle (3.0,2.0); % Near field border
        \draw[boxColor, very thick] (2.0,1.0) rectangle (2.5,1.5); % Active cluster border
    \end{scope}
  
      \begin{scope}[xshift=-60]
      % Adding labels
      \draw[-latex,thick] (0.0, 6.5) node[left]{Particle}        to [out=0,in=100]   (7.3, 5.3);
      \draw[-latex,thick] (0.0, 2.0) node[left, text width=1.6cm, text centered]{Evaluation point}  to [out=0,in=250]   (8.9, 3.6);
    \end{scope}
  
      \begin{scope}[xshift=390]
      % Adding labels
      \draw[thick] (0.0, 10.0) node[right]{Far-field}    to [out=180,in=300] (-3.0, 10.0);
      \draw[thick] (0.0, 6.5) node[right]{Near-field}   to [out=180,in=45] (-5.8, 5.3);
      \draw[thick, boxColor] (0.0, 2.0) node[right, text width=1.6cm, text centered]{Evaluation cluster}        to [out=180,in=300] (-6.5, 3.0);
    \end{scope}
    
  \end{scope}

\end{tikzpicture}
    \caption{The computational domain is hierarchically decomposed into areas that in turn are used to cluster the particles. Using the hierarchical decomposition, the near-field and far-field for a given target domain are identified. The hierarchy is used to find spatial relations between clusters at different levels. The combination of clusters at different levels are used to approximate the far-field interactions.}
    \label{fig:spatialDecomposition}
\end{figure*}

In a nutshell, the {\fmm} algorithm accelerates the evaluation of all the pairwise particle interactions by approximately evaluating them on clusters of particles. At the core of the algorithm, a hierarchical decomposition of space is used to group particles into clusters at different scales, as shown in Figure~\ref{fig:spatialDecomposition}. In the {\fmm}, the influence of a cluster of particles is approximately represented by a Multipole Expansion (\ME), which is an infinite series expansion. The {\ME} is truncated after a given number of terms, enabling the control of the accuracy of approximation. By using the {\ME}, it is possible to evaluate a distant cluster of sources at once and as a whole. The use of the hierarchical decomposition of space and the \ME s reduce the computational complexity of the algorithm to $\mathcal{O}(N\log N)$. To further reduce the computational complexity to $\mathcal{O}(N)$, the {\fmm} introduces the concept of Local Expansions (\LE s). An {\LE} is a series expansion that converges in a local domain. It is used to evaluate the \emph{aggregated effect of the {\ME}s of a group of clusters within the local domain}. By converting all the {\ME}s that represent the far-field into a single {\LE}, the effects due to interactions with particles in the far-field can be obtained in a single evaluation. Finally, to fully evaluate the domain, the contribution of the interactions with the particles in the near-field are computed directly.

\medskip

The {\fmm} can be organized into 5 stages:
\begin{enumerate}
\item \emph{Initialization}--- hierarchical decomposition of the space and generation of clusters of particles.
\item \emph{Upward sweep}--- creation of \ME s for all clusters.
\item \emph{Downward sweep}--- transformation of {\ME}s into {\LE}s, and aggregation of {\LE}s.
\item \emph{Calculation of the far-field}--- \LE\ evaluation.
\item \emph{Calculation of the near-field}--- direct evaluation with particles in the near-field.
\end{enumerate}

Of the stages enumerated above, the most computationally intensive stages are the \emph{downward sweep} and the \emph{calculation of the near-field}. In many experiments with a serial implementation of the \fmm, we observed that these two stages can take around $99\%$ of the overall computing time for $N$ in the order of 10 million, and around 95\% for $N$ in the order of 4 million. In a parallel implementation, some overheads noticeably appear which reduce the fraction of these stages with respect to the total runtime, although they still substantially dominate. Complete timing breakdowns of a parallel {\fmm} algorithm developed in our group are given in \cite{CruzKnepleyBarba2010}. In the present paper we will address the {\gpu} implementation of one of the two dominating stages of the overall algorithm: the downward sweep. The second one, the calculation of the near-field, is a pleasantly parallel problem and it has already been addressed in \cite{NylandHarrisPrins2007,BellemanEtal2008}.  We would like to note here also that a first publication of work on implementing the full \fmm\ algorithm for \gpu s was presented in \cite{GumerovDuraiswami2008}.

%!TEX root = CruzLaytonBarba2010.tex

\subsection{Fast Gauss transform}\label{fgt_background}

The fast Gauss transform ({\fgt}) was originally proposed by Greengard and Strain \cite{GreengardStrain1991}, as a specialization of the {\fmm}, for the case when the interaction function $\uKern$ from Equation~\eqref{eq:n_body} takes the form of a Gaussian kernel, as follows: 
\begin{equation}
	\label{eqn:dgt}
	G(y) = \sum_{i=1}^N q_i \, \exp\left( \frac{-\| x_i - y\|^2}{2\sigma^2} \right)
\end{equation}

\begin{figure*}
	\centering
	\includegraphics[width=0.7\textwidth]{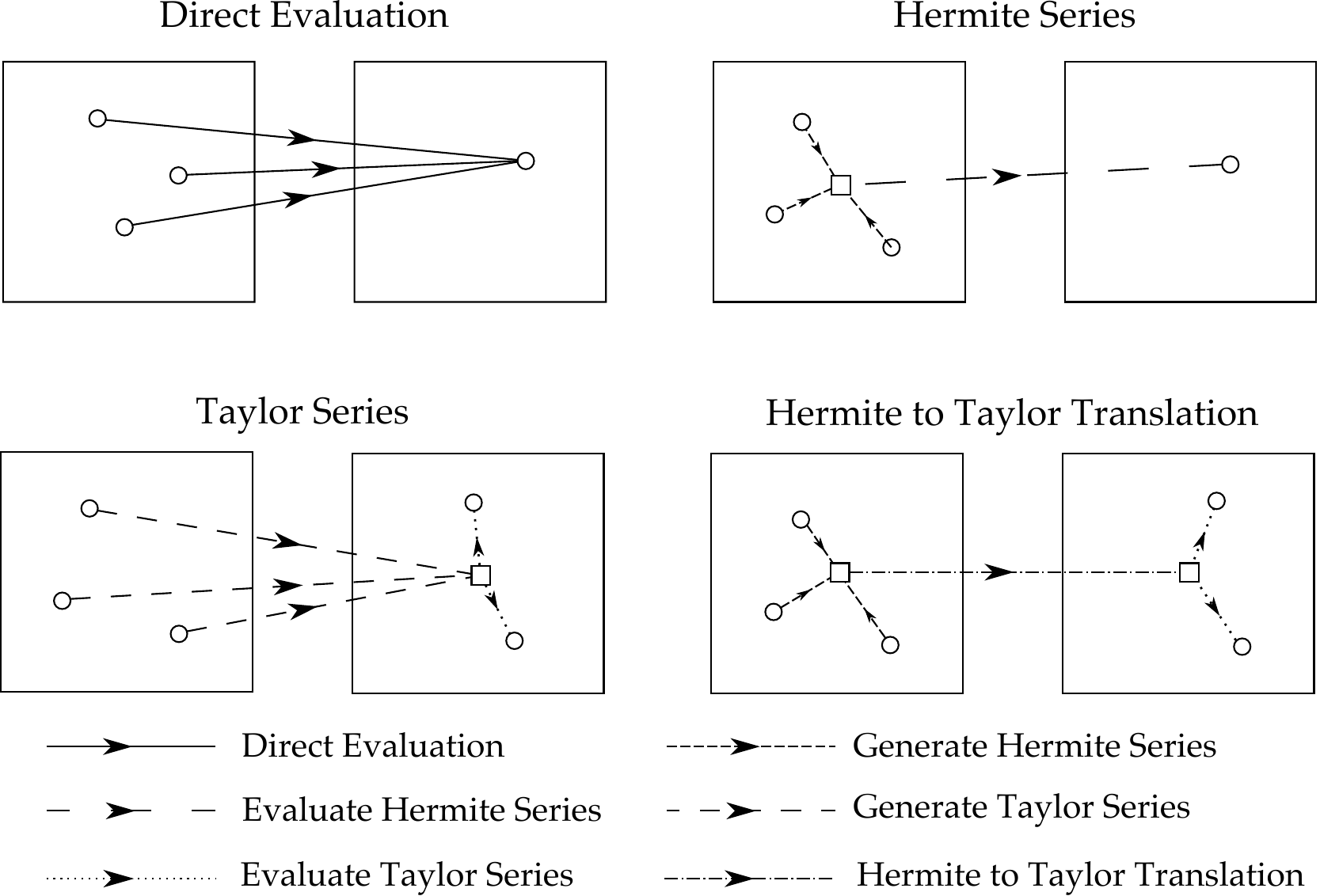}
	\label{fig:fgt_full}
	\caption{Illustration of the $4$ main interactions in the {\fgt} as previously enumerated. Source clusters are on the left, target clusters on the right. Arrows indicate direction of interaction. }
\end{figure*}

Similarly to the {\fmm}, the {\fgt} makes use of analytical approximations to reduce the computational complexity from {\bigOsq} to {\bigON}. The {\fgt} relies on two simple but powerful ideas: first, that Gaussians at a large distance from an evaluation point can safely be ignored with no detriment to accuracy; and second, that multiple individual Gaussians can be combined and approximated using a single series expansion. As with many other fast summation methods, such as the {\fmm}, the computational domain is decomposed in order to leverage these two ideas. In the particular case of the {\fgt}, a decomposition based on uniform box clusters is used, but the algorithm is agnostic towards the clustering scheme, a useful property in higher numbers of dimensions where simple boxes or cubes may be inefficient. Once the computational domain has been divided, it is necessary to choose what kind of approximation will be used to calculate the potential between any given pair of source and target clusters. There are four distinct options, chosen by work estimates and outlined below; they are illustrated graphically in Figure \ref{fig:fgt_full}.

\begin{enumerate}
	\item \emph{Direct evaluation}--- Use \eqref{eqn:dgt} for direct evaluations between individual source and target particles.
	\item \emph{Hermite series evaluation}--- Generate Hermite expansions from sources around \emph{source} centers and evaluate against individual target particles.
	\item \emph{Taylor series evaluation}--- Generate Taylor expansions from sources around \emph{target} centers and evaluate against target particles.
	\item \emph{Hermite to Taylor series translation}--- Generate Hermite series around \emph{source} centers, then translate these series to Taylor series around \emph{target} centers. Finally evaluate these Taylor series at target particles.
\end{enumerate}

We will briefly describe each of these methods in turn, and the situations in which they are used. For a more detailed description of the algorithm, we refer the reader to the original reference \cite{GreengardStrain1991}. The first form of interaction, the \emph{direct evaluation}, corresponds to applying the original sum of Equation \eqref{eqn:dgt}. However, rather than evaluating all source points in the computational domain against a given target sub-domain, only the sources within a local sub-domain are considered\,---\,by doing so, the entire sum does not devolve into {\bigOsq} complexity. For this reason, direct evaluation is only used where the number of both sources and targets is small.

In cases where there are a large number of source and target particles, one uses a more efficient way of calculating the potential than direct evaluation, based on series expansions. There exist three different interaction approximations that can be used for this purpose, these are: Hermite series \eqref{eqn:A_alpha}, Taylor series \eqref{eqn:B_beta}, or an approach that combines the use of both by translating a pre-existing Hermite series into a Taylor series around a different point \eqref{eqn:C_beta}. In practice, the approximation to be used depends on the computational efficiency of a given situation. For all series, the accuracy can be controlled by choosing a pair of multi-index variables, $\alpha$ and $\beta$, with $h_n(t) = e^{-t^2}H_n(t)$ and $H_n$ denoting the $n$th Hermite polynomial. Furthermore, we use the relation $H_{\alpha}(t) = H_{\alpha_1}(t) \cdots H_{\alpha_d}(t)$ with the $d$-dimensional multi-variate indices $\alpha$ and $\beta$. Other relevant properties of multi-variate indices used in the {\fgt} are as follows: $\alpha! = \alpha_1!\cdots\alpha_d!$, $|\alpha| = \alpha_1 + \cdots + \alpha_d$ and $t^{\alpha} = t^{\alpha_1}\cdots t^{\alpha_d}$.

\begin{eqnarray}
	\centering
	\label{eqn:A_alpha} A_{\alpha} & = & \frac{1}{\alpha !} \sum_{j=1}^{N_B} q_j \left ( \frac{x_j-s_B}{\sqrt{2}\sigma} \right )^{\alpha} \\
	\label{eqn:B_beta} B_{\beta} & = & \sum_{j=i}^{N_B} q_j\frac{(-1)^{|\beta|}}{\beta !} h_{\beta} \left ( \frac{x_j - t_C}{\sqrt{2}\sigma} \right ) \\
	\label{eqn:C_beta} C_{\beta} & = & \sum_{\alpha \geq 0}A_{\alpha}h_{\alpha+\beta}\left ( \frac{t_c-s_B}{\sqrt{2}\sigma}\right )
\end{eqnarray}

\noindent The first series \eqref{eqn:A_alpha} is used in the case where there are many source points but comparatively few targets. In this case, a Hermite series is created about a cluster center, $s_B$, with coefficients $A_\alpha$. 
The second alternative is a Taylor series about the center of the \emph{target} cluster, $t_C$. This is used when there are a large number of target points, but comparatively few sources. This ensures that time is not wasted generating expensive Hermite expansions when there are not enough sources to justify the cost.
Finally, when there is a large number of both sources and targets, one uses a combination of both types of series. First, a Hermite series is created about the center of the \emph{source} cluster as before, then that series is translated into a Taylor series about the center of the \emph{target} cluster, which can then be evaluated.

Once all the necessary series and translations have been created, they must be evaluated to give the final potential using either \eqref{eqn:hermite} for Hermite series, or \eqref{eqn:taylor} for Taylor expansions.

\begin{eqnarray}
	\centering
	\label{eqn:hermite} G(y) &=& \sum_{\alpha \geq 0} A_{\alpha}h_{\alpha}\left ( \frac{y-s_B}{\sqrt{2}\sigma} \right )\\
	\label{eqn:taylor} G(y) &=& \sum_{\beta \geq 0} B_{\beta} \left ( \frac{y-t_C}{\sqrt{2}\sigma} \right )^{\beta}
\end{eqnarray}

Using combinations of the different evaluation strategies one can approximately represent the potential field at the evaluation locations. All of the options for evaluating potentials that have been described above are combined to form a fast, efficient algorithm that runs with {\bigON} complexity. It is notable that all the formulae used are agnostic towards the number of dimensions of the Gaussian basis. Thus, the \fgt\ is particularly useful in fields where problems of this kind occur in high dimensions, as in options pricing \cite[]{BroadieYamamoto2003}, information theory \cite[]{HanETal2006}, image processing \cite[]{YangETal2003_2}, and others.

%% SECTION 3
\section{Motivation:  the opportunities of the new hardware architectures}\label{motivation}

%!TEX root = CruzLaytonBarba2010.tex
%%   hardware.tex

%% CHECK:
% Use SIMT model
% verify MAD definition

%%% SECTION
%\section{Background and motivation:  the opportunities of the new hardware architectures}} \label{hardware}

We focus on the {\cuda} architecture and programming framework from \NV. That said, similar hardware produced by other vendors (such as \textsc{amd} and \textsc{intel}) as well as different programming frameworks (such as {\opencl}), rely on the same technology paradigm of massively-parallel and throughput-optimized processors.  Therefore, the principles that we discuss here can be applied generally to current {\gpu} technologies, independent of vendors.

\subsection{The {\cuda} architecture and programming model}\label{gpu}

% CUDA
Developed by {\NV}, {\cuda} is  a parallel programming model, built for getting the most out of their {\gpu} hardware architecture. This model can involve hundreds of processing cores, and many thousands of individual threads running at any given time. This massive amount of parallelism forces one to focus on developing inherently parallel algorithms in order to get performance. On the other hand, the \NV\ {\gpu} architecture and {\cuda} programming model can mitigate some of the complications of parallel programming, many being handled by the hardware and transparent to the programmer.

% kernels & threads
In this programming model, a parallel program is executed as a \emph{kernel}. These kernels are simply programs written in the C language that will be executed by a single thread. It is generally the case that a kernel will be executed by thousands of simultaneous threads. Each of these threads is extremely lightweight, with no appreciable creation overhead. It also has a unique \textsc{id} number, with independent control flow and memory space. A further advantage is that active threads can be ``hot-swapped'' by the {\gpu}. Thus, when a thread is stalled due to a memory access, it can be swapped to another thread with no extra cost. This, combined with the fact that optimal efficiency is gained with tens of thousands of threads, allows us to hide the significant memory latency inherent in the architecture. When a kernel is executed, it is run on a number of threads that is explicitly defined by the user. On the {\gpu}, the threads are handled as groups of \emph{warps} (a 32-thread unit) and all the threads in a warp execute using a single-instruction/multiple-thread model (\textsc {simt}; see Table \ref{tab:Acronyms} for acronyms). Under the \textsc {simt} model, all threads will perform the same instruction and follow the same code execution path. Additionally, the \textsc {simt} model also allows the threads to branch in their execution but with a reduced execution performance.

% cooperation
To avoid unnecessary computations, the {\cuda} model allows both shared results and shared memory accesses. These combine to reduce both redundant computations and redundant memory accesses. This kind of cooperation, however, does not scale well to the thousands of concurrent threads that we have already established as the key {\cuda} methodology. The solution given by {\cuda} is to limit co-operation to small groups of threads, called \emph{thread blocks}.  These blocks are scalable, and allow synchronization between their component threads. A thread block is assigned and executed by a single streaming-multiprocessor (\textsc{sm}). One important characteristic of this model, is that it does not allow for communication between different blocks. This grouping is the model used by {\cuda} to scale on the hardware with different numbers of available \textsc{sm}s. The more powerful {\gpu}s will have more \textsc{sm}s available and can execute more thread blocks simultaneously.

% -------------------------------------------- Acronyms Table
\begin{table*}
	\centering\vspace{0mm}
	\begin{tabular} { |l  l| }
	\hline
	Acronym & Meaning \\ \hline
\rowcolor[gray]{0.9} SPMD & Single Program, Multiple Data \\
\rowcolor[gray]{1.0} SPMT & Single Program, Multiple Thread \\
\rowcolor[gray]{0.9} TPC & Texture Processor Clusters \\
\rowcolor[gray]{1.0} SM & Streaming Multi-Processors \\
\rowcolor[gray]{0.9} SP & Streaming Processors \\
\rowcolor[gray]{1.0} SIMT & Single-Instruction, Multiple Thread \\
\rowcolor[gray]{0.9} SFU & Special Function Units \\
\rowcolor[gray]{1.0} FPU & Floating-Point Units \\ \hline
	\end{tabular}
	\label{tab:Acronyms}
	\caption{Acronyms for \gpu\ architecture.}
\end{table*}

%  SUBSECTION
\subsection{Key features of the {\gpu} computing technology}\label{hardware}   % ------------  GPU

The essential hardware features of  \gpu s are very different from \cpu s: they have many times more processor cores, and a totally different memory system.  For example, the main characteristics of the {\NV} {\gpu} chip used in this study\,---the GT200, which is used in both the commodity GeForce GTX $2xx$ video cards and the Tesla C1060 computing processor---\,are the following\footnote{Most of the results in this paper, and all of the development, were made on Tesla-series GT200 \gpu s. During revision, we gained access to the newest generation \gpu\ (Fermi architecture) and thus we were able to add results obtained on this chip. The discussion in the text, however, does not emphasize Fermi.}:

\vspace{\hup}

\begin{itemize}
\item[$\triangleright$] 240 `streaming processor' cores per {\gpu} chip, clocked at 1.296 GHz\vspace{\hup}
\item[$\triangleright$] cores grouped into `streaming multi-processors', with 8 cores each\vspace{\hup}
\item[$\triangleright$] each multi-processor has a \emph{shared memory} of 16 kB 
\end{itemize}

\noindent Note the large number of processor cores, and the small amount of \emph{shared memory} of the {\gpu}.  The {\gpu} is specialized, highly-parallel hardware, appropriate for computationally-intensive tasks.  To be specific, the {\gpu} is a \emph{streaming, pipelined architecture}.  It is `pipelined' because it is built for processing each of many objects in the same way (which is the situation in graphics rendering).  The architecture is optimized for \emph{homogeneous units of work}.  It is `streaming' because it allows a large number of simultaneous computational units, with synchronization and communication among these units being provided by the hardware.  These fundamental architectural attributes make the {\gpu} exceptionally suited to perform computations that have a high level of data parallelism.

Now, let us consider the performance capability of the {\gpu} described above.  To satisfy the reader of the correctness of the marketed performance numbers for this chip, we have to go into a little more detail about the architecture than we would like to.  But, if the reader will bear with us, this will prove revealing.

\begin{figure*}
	\centering
	\vspace{-2mm}\includegraphics[width=0.55\textwidth]{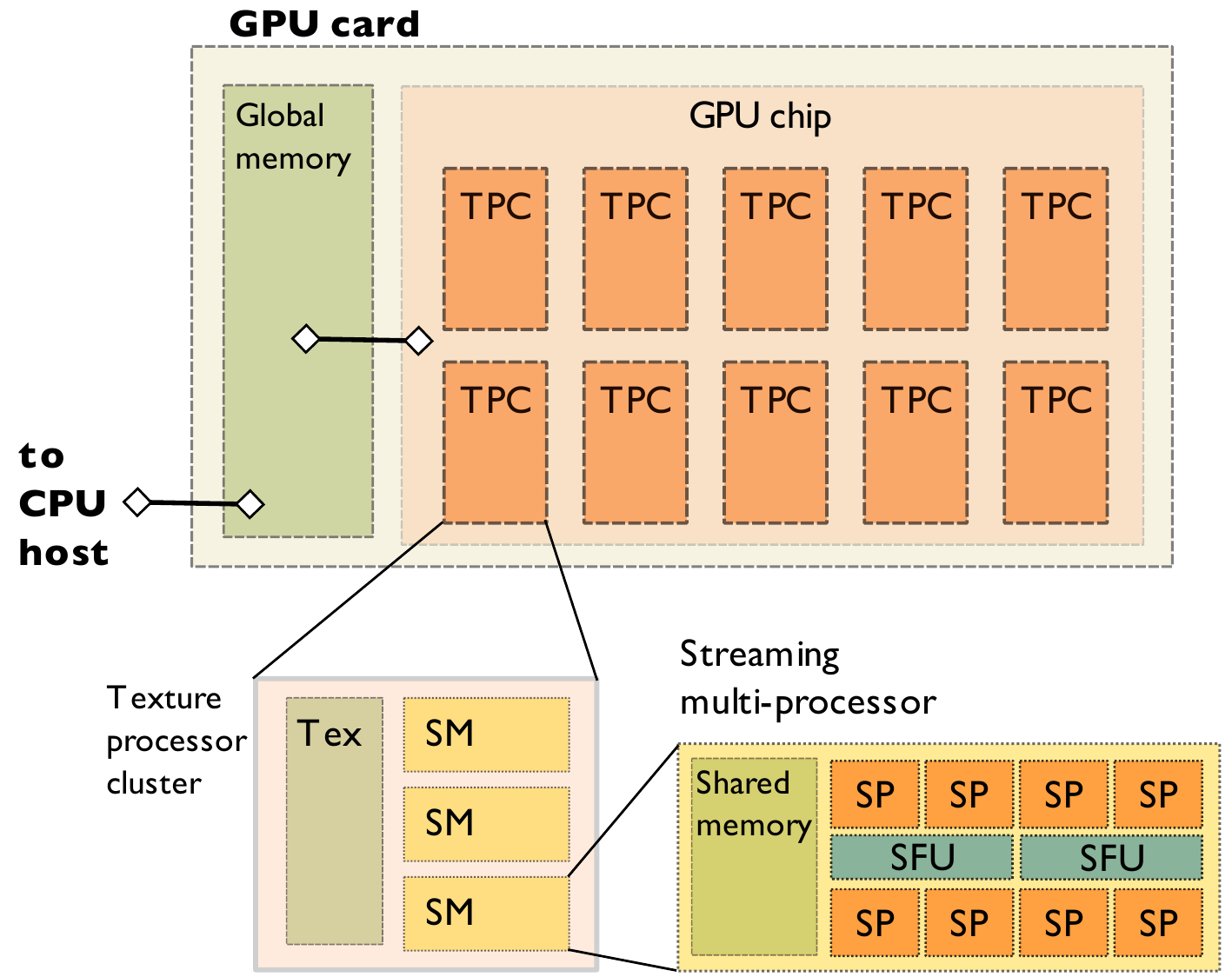}
	\caption{Details of GT200 \gpu\ architecture.}
	\label{fig:GPUcard}
\end{figure*}

The GT200 chip really has 10 processor clusters, or {\tpc}s (see Figure \ref{fig:GPUcard} for a sketch of the {\gpu} hardware  architecture).  Each of the {\tpc}s has 3  \textsc{spmt} computing cores  (\textsc{sm}), consisting of streaming processors (\textsc{sp}) and special-function units (for transcendentals, \emph{etc}.).  An \textsc{sm} can issue, in each cycle, instructions to \emph{either}:

\begin{itemize}
\item[$\triangleright$] 8 single-precision \textsc{fpu}s (\emph {stream processors})\vspace{\hup}
\item[$\triangleright$] 1 double-precision \textsc{fpu}\vspace{\hup}
\item[$\triangleright$] a branch unit that manages \textsc{simt} execution
\end{itemize}
\noindent A \emph{streaming processor} can perform a single precision multiplication and an addition (\textsc{mad}) in each cycle (2 flop/cycle), thus contributing the following to the performance:

\vspace{\hup}
% ------------------------------------------- performance Port 0
\begin{eqnarray*}
1.296 \text{ GHz } \times 10 \text{ \textsc{tpc} } \times 3 \text{ \textsc{sm} } \times 8 \text{ \textsc{sp} } \times 2 \text{ flop/cycle} &\\= 622.08 \text{ Gflop/s}.&
\end{eqnarray*}
The special function units, meanwhile, can compute 4 floating point 
operations per cycle in a vector instruction, and thus contribute the
following to the performance:

\vspace{\hup}
% ------------------------------------------- performance Port 1
\begin{eqnarray*}
1.296 \text{ GHz } \times 10 \text{ \textsc{tpc} } \times 3 \text{ \textsc{sm} } \times  2 \text{ \textsc{sfu} } \times 1 \times  4 \text{ flop/cycle} &\\= 311.04 \text{ Gflop/s}.&
\end{eqnarray*}

\noindent The sum of the throughput that can be obtained from the streaming processors and the special function units is therefore 933 Gflop/s, the advertised single-precision peak performance of the \gpu.  It must be noted that this peak assumes that instructions are constantly being co-issued to the \textsc{sp}s and \textsc{sfu}s.  When no special functions are required in an algorithm, it is possible to use the special-function unit to execute instead single-precision multiplies, thus in theory the peak throughput is available. However, arguments still arise when various researchers meet to discuss their results as percentage of ``peak''; some consider the practical peak to be 622 Gflop/s, for example.  In double precision, on the other hand, there is only one \textsc{fpu} available per \textsc{sm} and thus the peak performance of the {\gpu} is: 1.296 MHz $\times$ 30 \textsc{sm} $\times$ 1 \textsc{fpu} $= 78$ Gflop/s (the Fermi architecture improved on this considerably). 

While this analysis focuses on floating point performance, we are just as interested in overall performance of the {\gpu}, also taking into account integer operations. Thus, we wish to take our analysis based of Gflop/s and try to draw some conclusions about the general performance in Operations/s (Gop/s). From \cite{cuda-guide3.0}, we can obtain the throughput of different arithmetic functions for different datatypes for compute capabilities $1.3$ and $2.0$. For $1.3$-capable cards (such as the Tesla C1060), we see that integer multiply and mutiply-add are listed as `multiple instructions', while for $2.0$ compute capability (such as the Fermi C2050), the throughput is identical for that of equivalent floating point operations. From this, we conclude that for the C1060 peak performance in Gop/s is slightly lower than the value determined above in Gflop/s, but for the C2050 cards, performance in floating point and integers should be equivalent.

% -------------------------------------------- GPU card
   % ------------ 

Clearly, there will be many challenges in a particular application for extracting the promised performance out of the {\gpu}.  The throughput capacity calculated above refers to the \emph{maximum} number of instructions that can be issued on the chip. Moreover, the \emph{actual} throughput will also depend on issues related to the access to memory, in particular the high-latency accessess to global memory from the {\gpu} chip (where memory transactions range between 400 and 600 clock cycles), and the use of shared memory in the streaming multi-processors to mitigate the high-latency effects.

 For the algorithms of interest in this work, we have developed formulations that demonstrate the power of \emph{heterogeneous computing} using {\gpu}s. The subsequent section will discuss the work flow that was followed for the formulation of the algorithmic kernels in such a way as to maximize performance in the architecture.

%% SECTION 4
\section{Formulation of the algorithmic kernels for the GPU and optimization}\label{gpuformulation}

%!TEX root = CruzLaytonBarba2010.tex

%% SECTION: Formulation of the algorithms for the GPU and optimization

% hardware metrics
When formulating algorithms for the {\gpu}, one wants to have the means to analyze their performance and efficiency in terms of resource utilization for the target architecture. By performing such analysis, we can predict the maximum theoretical performance of the algorithm on the given hardware, and use this information as a guide during the design and optimization of the algorithm.

Before any analysis can take place, we need to build a simplified model of the {\gpu} that characterizes all the important features of the architecture. From our introduction in section \S\ref{hardware}, a current-generation {\gpu} can be described as a \emph{massively parallel architecture} that has been \emph{optimized for high throughput}. In our simplified model, we consider each {\gpu} to have hundreds of very simple processors (processing elements) that can execute concurrently. We will assume that each processing elements is capable of computing the following operations, in single precision and at a constant cost: arithmetic operations ($+, -, \times, \div, \mathrm{transcendental~functions}$), and conditional statements. Regarding the data movement operations (load, store, and copy), we will not take into account all the details of the {\gpu} memory hierarchy but use a simplified model where we distinguish between off-chip memory (global memory) and on-chip memory (registers and shared memory)\,---the main considerations are the limited bandwidth available and limited capacity, respectively.  

Using the simplified {\gpu} model, we can estimate the operation complexity and data accesses of an algorithm. We use this information  to evaluate the algorithmic performance against the hardware theoretical \emph{throughput} and \emph{bandwidth}. 

\begin{itemize}
	\item[$\triangleright$]  \emph{Throughput (measured in op/s):} 
	The average number of operations per second that can be executed by the {\gpu}. A large number of op/s implies a high level of theoretical peak performance. Traditionally, this number is presented as Giga-op/s, abbreviated as Gop/s. Giga-flop/s or Gflop/s are also commonly presented, but using op/s gives a better idea of how close to the peak performance of the {\gpu} an algorithm obtains.
\vspace{\hup}
	\item[$\triangleright$]  \emph{Bandwidth (measured in Giga-Bytes/sec, GB/s):} 
	The rate at which data is transferred between the memory and the processor. Bandwidth measures how much data is being read and written from global memory in a unit time. In practice, many applications are bandwidth-limited, therefore it is important to use the bandwidth as efficiently as possible.
\end{itemize}

% Algorithmic properties
 In addition to our {\gpu} model, we define four algorithmic properties that map to different features of the {\gpu} paradigm.  These properties are then used as guidelines when rethinking an algorithm:

\begin{itemize}
\item[$\triangleright$] \emph{Computational intensity}: ratio of operations performed to memory accesses in a unit of time. A compute-intensive task will have an operations-to-memory access ratio larger than the one given by the theoretical performance of the {\gpu}.  Highly compute-intensive tasks benefit from the high throughput of the {\gpu}, while low compute-intensive tasks benefit from the large memory bandwidth of the {\gpu}.
\vspace{\hup}
\item[$\triangleright$] \emph{Concurrency}: parts of the algorithm that \emph{may} be executed simultaneously. There are many levels of concurrency that range from coarse-grained concurrency to fine-grained concurrency. The latest {\gpu}s can benefit from multiple levels of concurrency. For instance, when using {\cuda}-enabled devices, one is able to explicitly identify concurrency at different levels,  coarse- to fine-grained: \cuda\ kernel, thread-block, and thread calculations.
\vspace{\hup}
\item[$\triangleright$] \emph{Homogeneity}: degree at which concurrent computations are the same, independently of their input. Concurrent tasks that present a high degree of homogeneity can take full advantage of the `single-instruction /multiple-thread' \textsc {simt} model  of the {\gpu} architecture.
\vspace{\hup}
\item[$\triangleright$] \emph{Data-locality}: the way in which the physically stored data is accessed by an algorithm. The role of data locality is twofold: a high degree of spatial data locality can result in more efficient data accesses, \emph{e.g.}, when data that is spatially adjacent can be accessed by thread collaboration, resulting in more efficient memory transactions. A high degree of temporal data locality avoids redundant data transfers of data values that are frequently accessed within small time windows.
\end{itemize}

An ideal algorithm for the \gpu\ would be computationally intensive and have a high degree of concurrency. Concurrent computations would be highly homogeneous and show a large amount of data locality. However, in practice these properties are difficult to measure and as such do not allow one to effectively quantify the expected performance of the algorithm under study.

%!TEX root = CruzLaytonBarba2010.tex
%%   fmm_gpu.tex

%%%  SUBSECTION
\subsection{Fast multipole method for the {\gpu}}

We will now concentrate on the algorithmic redesign of the {\fmm} for {\gpu} computing, whereas a more in-depth discussion of the implementation details takes place in \S\ref{implementation}. As such, we start by noticing that the most time-consuming stages of the {\fmm} algorithm are: the near-field calculations, and the downward sweep. As we observed in \S\ref{subsec:fmm_back}, these stages can take $95-99\%$ of the runtime in a serial implementation of the algorithm, and continue to dominate in parallel implementations despite parallel overheads.

Let us discuss in more detail the computations performed by these two stages.  In the case of the near-field calculations, the problem is to compute the mutual interactions of particles in many small- to medium-size domains, with each domain containing at most a few hundred particles. Therefore, the near-field calculation problem can be reduced to efficiently implementing the direct particle interactions for each domain, and as such, this is a special case of the $N$-body problem, where interactions are computed only with a subset of the $N$ particles. Furthermore, the near-field calculation retains the distinctive features of being computationally intensive and highly parallel. Therefore, this problem can be efficiently mapped to the \gpu\ architecture, achieving high performance, as reported in~\cite{NylandHarrisPrins2007,BellemanEtal2008}.
We note that, as pointed out by \cite{NylandHarrisPrins2007}, a balanced \fmm\ execution in the \gpu\ will perform the near-field direct evaluation with a larger set of particles per box than in the {\cpu} case. This difference is due to the higher efficiency achieved by the {\gpu} in the near-field calculations, as compared to the far-field calculations. Consequently, the optimal workload distribution between these two stages is different in the {\gpu} and {\cpu} executions.

In the downward sweep, the computations are dominated by the transformation of multipole expansions into local expansions ({\ML} step). Tens of thousands of concurrent {\ML} transformations are computed in the {\fmm} algorithm, this being the reason for such a high computational intensity. Each of these transformations can be expressed in the form of a matrix-vector multiplication, where each matrix is dense, of size ($p \times p$), where $p$ corresponds to the number of terms of a truncated {\ME}. 
The algorithmic steps of the downward sweep are: 
\begin{enumerate}
\item For every node of the hierarchical tree, we obtain the list of clusters that belong to the same level, effectively computing the interaction list of each node. 
\item Then for every node (or evaluation cluster) in the tree, the multipole expansions of each cluster in its interaction list are transformed into a local expansion centered at the evaluation cluster.
\item Finally, the cluster's local expansion is obtained by aggregating the local expansions obtained from the {\ML} transformation of all the clusters in the interaction list, this step is referred to as the \emph{reduction}.
\end{enumerate}

We present the following example to illustrate the parallelism and computational intensity of the downward sweep stage. Consider a two-dimensional case ($d = 2$) where the computational domain is hierarchically decomposed by a uniform tree (a quadtree) with $l=5$ levels of spatial refinement. Such a tree will approximately have $4^{l}$ nodes, where each node will transform on average $27$ multipole expansions (more precisely, this number corresponds to the size of the node's interaction list). Therefore, the approximate number of {\ML} transformations computed are $4^{5} \times 27 = 27,648$. For deeper trees, more transformations would be needed. 

In the original algorithm for the {\ML} transformation, the multipole expansion of a given cluster $j$, with center $x_j$ and multipole coefficients $m_j$, is transformed into a local expansion for cluster $i$ with center at $x_i$. The result of the transformation will give the local expansion coefficients $l_i$. In the $2$-dimensional case, both the coordinates and expansion terms are usually represented by complex numbers. The {\ML} transformation in its matrix-vector multiplication form has a transformation matrix $A$, with $p\times p$ complex elements, $a_{nk}$. These elements can be obtained using the following expression:
\begin{equation}\label{eq:M2Lmat}
a_{nk} = (-1)^{n} \left(
\begin{array}{c}
  n+k      \\
   k   
\end{array}
\right)
{\left( x_{i} - x_{j} \right)}^{-k-n-1}
\end{equation} 

As previously noted, the {\ML} stage is highly parallel and computationally intensive. However, in order to obtain a high-performance implementation for the {\gpu}, the {\ML} stage needs to be reformulated.

%%% Reformulation of the {\ML} algorithm
\subsubsection{Reformulation of the {\ML} problem}

As a first approach,  one can consider distributing the mat-vec operations across threads.  That is, each {\cuda}-thread would perform one such operation.  Suppose that the evaluation requires a truncation parameter $p=12$, as required for high-accuracy simulations; see for example \cite{CruzBarba2009}. In that case, each matrix has a size of $2p^2=288$, requiring 2304 bytes of storage.  Thus, in the {\gpu} shared memory of 16~kB one could fit a maximum of $6$ such matrices, which implies that a maximum of $6$ threads could be run in one multi-processor at a time. Clearly this number of threads is much too small\,---as discussed in \S\ref{hardware}, more than one hundred threads per {\sm} are needed in order to hide memory transfer latencies. Alternatively to storing the transformation matrix in shared memory, it is feasible to use Equation \eqref{eq:M2Lmat} to compute the matrix elements as the matrix-vector multiplication takes place (also known as matrix-free format). 
%
% -------------------------------------------- Figure M2L versions
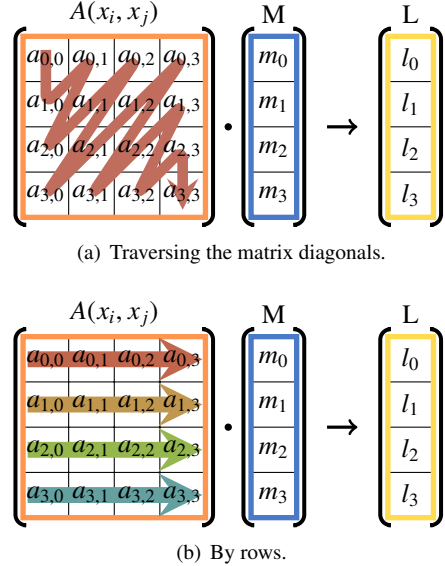
\begin{figure}
	\centering
	\subfigure[Traversing the matrix diagonals.]
		{\label{fig:m2la} %% Driagram of the M2L translations
\begin{tikzpicture}

% Define colors to use
\definecolor{meColor}{rgb}{0.29, 0.48, 0.78}
\definecolor{leColor}{rgb}{1.0,0.86,0.31}
\definecolor{m2lColor}{rgb}{1.0,0.59,0.31}
\definecolor{threadColor}{rgb}{0.65,0.18,0.10}

% Efficient Matrix vector multiply
\begin{scope}[scale=0.40]
    % Draw labels
    \draw (3,6.75) node {$A(x_i, x_j)$};
    \draw (8.25,6.75) node {M};
    \draw (12.75,6.75) node {L};
    % Fill matrix and vectors
    \draw[step=1.5,black] (0,0) grid (6,6); % Matrix
    \draw[step=1.5,black] (7.5,0) grid (9,6); % ME
    \draw[step=1.5,black] (12,0) grid (13.5,6); % LE
    % Color matrix and vector
    \draw[m2lColor, line width=2pt] (0,0) rectangle (6,6); % Matrix
    \draw[meColor, line width=2pt] (7.5,0) rectangle  (9,6); % ME
    \draw[->,black, very thick] (10,3) -- (11,3);
    \draw[leColor, line width=2pt] (12,0) rectangle (13.5,6); % LE
    \fill[black] (6.75,3) circle (0.12); % multiply
    % Draw Brackets
    % matrix
    \draw[black, line width=1.5pt, rounded corners]  (0, 6.25) -- ++(-0.25,0)  -- ++(0,-6.5) -- ++(0.25,0); % left
    \draw[black, line width=1.5pt, rounded corners]  (6, 6.25) -- ++(0.25,0)  -- ++(0,-6.5) -- ++(-0.25,0); % right
    % ME
    \draw[black, line width=1.5pt, rounded corners]  (7.5, 6.25) -- ++(-0.25,0)  -- ++(0,-6.5) -- ++(0.25,0); % left
    \draw[black, line width=1.5pt, rounded corners]  (9, 6.25) -- ++(0.25,0)  -- ++(0,-6.5) -- ++(-0.25,0); % right
    %LE
    \draw[black, line width=1.5pt, rounded corners]  (12, 6.25) -- ++(-0.25,0)  -- ++(0,-6.5) -- ++(0.25,0); % left
    \draw[black, line width=1.5pt, rounded corners]  (13.5, 6.25) -- ++(0.25,0)  -- ++(0,-6.5) -- ++(-0.25,0); % right
    % Draw computation path
    \draw[>=stealth,->, threadColor!70, solid, line width=4pt] {[rounded corners] 
          (0.75, 5.55) -- (0.75, 3.75)
      -- (2.25, 5.25) -- (0.75, 2.25)
      -- (3.75, 5.25) -- (0.75, 0.75)
      -- (5.25, 5.25) -- (2.25, 0.75)
      -- (5.25, 3.75) -- (3.75, 0.75)
      -- (5.25, 2.25) -- (5.25, 0.25)};
    % Matrix elements
    \draw (0.75, 0.75) node{$a_{3,0}$};%
    \draw (2.25, 0.75) node{$a_{3,1}$};
    \draw (3.75, 0.75) node{$a_{3,2}$};
    \draw (5.25, 0.75) node{$a_{3,3}$};
    \draw (0.75, 2.25) node{$a_{2,0}$};%
    \draw (2.25, 2.25) node{$a_{2,1}$};
    \draw (3.75, 2.25) node{$a_{2,2}$};
    \draw (5.25, 2.25) node{$a_{2,3}$};
    \draw (0.75, 3.75) node{$a_{1,0}$};%
    \draw (2.25, 3.75) node{$a_{1,1}$};
    \draw (3.75, 3.75) node{$a_{1,2}$};
    \draw (5.25, 3.75) node{$a_{1,3}$};
    \draw (0.75, 5.25) node{$a_{0,0}$};%
    \draw (2.25, 5.25) node{$a_{0,1}$};
    \draw (3.75, 5.25) node{$a_{0,2}$};
    \draw (5.25, 5.25) node{$a_{0,3}$};
    % ME elements
    \draw (8.25, 5.25) node{$m_{0}$};%
    \draw (8.25, 3.75) node{$m_{1}$};
    \draw (8.25, 2.25) node{$m_{2}$};
    \draw (8.25, 0.75) node{$m_{3}$};
    % ME elements
    \draw (12.75, 5.25) node{$l_{0}$};%
    \draw (12.75, 3.75) node{$l_{1}$};
    \draw (12.75, 2.25) node{$l_{2}$};
    \draw (12.75, 0.75) node{$l_{3}$};
\end{scope}
\end{tikzpicture}}
	\subfigure[By rows.]
		{\label{fig:m2lb} %% Driagram of the M2L translations
\begin{tikzpicture}

% Define colors to use
\definecolor{meColor}{rgb}{0.29, 0.48, 0.78}
\definecolor{leColor}{rgb}{1.0,0.86,0.31}
\definecolor{m2lColor}{rgb}{1.0,0.59,0.31}
\definecolor{thread0Color}{rgb}{0.65,0.14,0.00}
\definecolor{thread1Color}{rgb}{0.65,0.42,0.00}
\definecolor{thread2Color}{rgb}{0.40,0.60,0.00}
\definecolor{thread3Color}{rgb}{0.13,0.47,0.46}

% Efficient Matrix vector multiply
\begin{scope}[scale=0.40]
    % Draw labels
    \draw (3,6.75) node {$A(x_i, x_j)$};
    \draw (8.25,6.75) node {M};
    \draw (12.75,6.75) node {L};
    % Fill matrix and vectors
    \draw[step=1.5,black] (0,0) grid (6,6); % Matrix
    \draw[step=1.5,black] (7.5,0) grid (9,6); % ME
    \draw[step=1.5,black] (12,0) grid (13.5,6); % LE
    % Color matrix and vector
    \draw[m2lColor, line width=2pt] (0,0) rectangle (6,6); % Matrix
    \draw[meColor, line width=2pt] (7.5,0) rectangle  (9,6); % ME
    \draw[->,black, very thick] (10,3) -- (11,3);
    \draw[leColor, line width=2pt] (12,0) rectangle (13.5,6); % LE
    \fill[black] (6.75,3) circle (0.12); % multiply
    % Draw Brackets
    % matrix
    \draw[black, line width=1.5pt, rounded corners]  (0, 6.25) -- ++(-0.25,0)  -- ++(0,-6.5) -- ++(0.25,0); % left
    \draw[black, line width=1.5pt, rounded corners]  (6, 6.25) -- ++(0.25,0)  -- ++(0,-6.5) -- ++(-0.25,0); % right
    % ME
    \draw[black, line width=1.5pt, rounded corners]  (7.5, 6.25) -- ++(-0.25,0)  -- ++(0,-6.5) -- ++(0.25,0); % left
    \draw[black, line width=1.5pt, rounded corners]  (9, 6.25) -- ++(0.25,0)  -- ++(0,-6.5) -- ++(-0.25,0); % right
    %LE
    \draw[black, line width=1.5pt, rounded corners]  (12, 6.25) -- ++(-0.25,0)  -- ++(0,-6.5) -- ++(0.25,0); % left
    \draw[black, line width=1.5pt, rounded corners]  (13.5, 6.25) -- ++(0.25,0)  -- ++(0,-6.5) -- ++(-0.25,0); % right
    % Draw Parallel Computation path
    \draw[>=stealth,->, thread0Color!70, solid, line width=6pt] (0.15, 5.25) -- (5.90, 5.25); % thread 0
    \draw[>=stealth,->, thread1Color!70, solid, line width=6pt] (0.15, 3.75) -- (5.90, 3.75); % thread 1
    \draw[>=stealth,->, thread2Color!70, solid, line width=6pt] (0.15, 2.25) -- (5.90, 2.25); % thread 2
    \draw[>=stealth,->, thread3Color!70, solid, line width=6pt] (0.15, 0.75) -- (5.90, 0.75); % thread 3
    % Matrix elements
    \draw (0.75, 0.75) node{$a_{3,0}$};%
    \draw (2.25, 0.75) node{$a_{3,1}$};
    \draw (3.75, 0.75) node{$a_{3,2}$};
    \draw (5.25, 0.75) node{$a_{3,3}$};
    \draw (0.75, 2.25) node{$a_{2,0}$};%
    \draw (2.25, 2.25) node{$a_{2,1}$};
    \draw (3.75, 2.25) node{$a_{2,2}$};
    \draw (5.25, 2.25) node{$a_{2,3}$};
    \draw (0.75, 3.75) node{$a_{1,0}$};%
    \draw (2.25, 3.75) node{$a_{1,1}$};
    \draw (3.75, 3.75) node{$a_{1,2}$};
    \draw (5.25, 3.75) node{$a_{1,3}$};
    \draw (0.75, 5.25) node{$a_{0,0}$};%
    \draw (2.25, 5.25) node{$a_{0,1}$};
    \draw (3.75, 5.25) node{$a_{0,2}$};
    \draw (5.25, 5.25) node{$a_{0,3}$};
    % ME elements
    \draw (8.25, 5.25) node{$m_{0}$};%
    \draw (8.25, 3.75) node{$m_{1}$};
    \draw (8.25, 2.25) node{$m_{2}$};
    \draw (8.25, 0.75) node{$m_{3}$};
    % ME elements
    \draw (12.75, 5.25) node{$l_{0}$};%
    \draw (12.75, 3.75) node{$l_{1}$};
    \draw (12.75, 2.25) node{$l_{2}$};
    \draw (12.75, 0.75) node{$l_{3}$};
\end{scope}
\end{tikzpicture}}
	\caption{Sketch of different strategies for \ML\ computation.}
	\label{fig:M2Lversions}
\end{figure}   % ------------ 

When performing the matrix-vector product in matrix-free format, one can ``traverse'' the matrix such that the matrix terms are efficiently calculated by reusing the computations performed to obtain the previous terms. Therefore, one can reuse a part of both the binomial and the power terms by traversing the diagonals of the matrix, as sketched in Figure \ref{fig:m2la}.  This represents a strategy for matrix creation and multiplication that is efficient from the point of view of operations that it performs, in both the {\gpu} and {\cpu} architectures.  Moreover, for the {\gpu}, by assigning the computations of one matrix-vector multiplication to only one {\cuda}-thread, thread synchronization is not required. Consider now that each thread block performs the transformations of only one {\ME} for all the interaction list, accounting for an estimated $27$ {\ML} transformations. During computation, the results of the transformation are stored in shared memory. With a maximum  of $8$ concurrent thread blocks per multi-processor in the GT200 \gpu, we have a maximum of $27\times 8=216$ active threads running on the multiprocessor.  This \NV\ {\gpu} allows a maximum of 512 threads in each thread block and 1024 active threads per {\sm}~\cite[][p.\,8]{cuda-guide}, and thus the approach just described could achieve less than one third of the maximum thread capacity of each {\sm}, without considering the register usage per thread. 
When the translation of an {\ME} is finished, one needs to move the result from shared memory to global memory, in the form of one {\LE} consisting of $2p$ floats. Hence, transforming one {\ME} for all the interaction list results in the movement of $27\times 2p$ floats. 
For instance, translating one {\ME} of $p=12$ results in 648 floats that need to be stored in global memory, which at the hardware level results in  a minimum of twenty 128-byte and one 32-byte memory transactions (see \S\ref{subsec:mem_management} for a more complete discussion on memory management). When transferring the results to global memory, the memory transactions are executed by the thread-block one after the other, resulting in a stage where only memory transactions take place. A considerable disadvantage of such a memory-intensive stage is that multithreading is less effective in hiding the  memory latency when the number of active threads is small. 
Nevertheless, our implementation using this approach sustained 20 Gop/s and was able to perform $2.5$ million translations per second, using a single C1060 {\tesla} card. %In comparison, our single-threaded {\cpu} implementation only achieved $140,000$  translations per second running on a 2.5 GHz Intel Core 2 Duo. 

\medskip

%%%%% Loop known number of iterations %%%%%
\begin{figure*}%[htbp]
\begin{center}
	\begin{lstlisting}
	int k = 12;
	int n = threadIdx.x;
	    
	float ac, bd;
	tkn1_r = t_r;
	tkn1_i = t_i;
	    
	for (m = 1; m < k+n+1; m++)  // tkn1 = tkn1 * t
	{
	    ac      = tkn1_r;
	    bd      = tkn1_i;
	    tkn1_r = ac * t_r - bd * t_i;
	    tkn1_i = ac * t_i + bd * t_r;
	}

	\end{lstlisting}
\caption{Code snippet for computing the complex power $t^{k+n+1}$, with $t = t_\Re + i * t_\Im$. This code shows a for-loop with a variable number of iterations.}
\label{code:loop_divergence}
\end{center}
\end{figure*}

%%%%% Loop known number of iterations %%%%%
\begin{figure*}%[htbp]
\begin{center}
	\begin{lstlisting}
	#define P 12
	
	int m;
	float ac, bd;
	float tn1_r = t_r;
	float tn1_i = t_i;
	
	#pragma unroll
	for (int m = 1; m < P; m++)
	{
	    if (threadIdx.x >= m) // tn1 = tn1 * t
	    {
	        ac   = tn1_r;
	        bd   = tn1_i;
	        tn1_r = ac * t_r - bd * t_i;
	        tn1_i = ac * t_i + bd * t_r;
	    }
	}

	\end{lstlisting}
\caption{Code snippet for computing the complex power $t^{n+1}$, with $t = t_\Re + i * t_\Im$. This code shows a for-loop where the number of iteration has been fixed at compile time to $P-1$, and it uses a conditional for computing only the relevant power to each thread.}
\label{code:loop_known_num_iter}
\end{center}
\end{figure*}

The performance of the implementation just described is very much below the theoretical peak for both throughput and bandwidth of a C1060 {\tesla} card. An assessment of the {\cuda} kernel implementation revealed that its main limitation was the inefficiency of the memory transfers to and from global memory. This was due to two problems: first, many of the memory transfers were non-coalesced (i.e. adjacent threads reading adjacent memory values, see \S\ref{subsec:mem_management} for a full explanation), thus the \emph{effective bandwidth} used by the kernel was very low when compared against the theoretical performance of the card; and second, the kernel did not have enough active threads working per {\sm} to effectively hide the latency of the memory transfers, making the efficiency of the kernel implementation suffer. These two problems can be tracked down to a flaw in the design of our algorithm: using only one {\cuda} thread to perform a single translation.
The resulting kernel was resource-intensive since the diagonally traversing matrix-free multiplication required too many variables to store the state of the computations and to control the flow of the program, thus limiting the total number of active threads per {\sm}.
Additionally, having only one thread to perform each translation made it difficult to use multiple threads to perform collaborative memory transactions. 

Desirable features of a more optimized kernel would be for it to be: (\emph{i}) \emph{simpler}, requiring less variables to store the state of its execution, thus allowing more threads to be active at a time; (\emph{ii}) \emph{more parallel}, allowing the use of more concurrent threads, and to use coordinated collaboration between threads for performing memory transfers. 

Now, let us examine again Equation~\eqref{eq:M2Lmat}. A more parallel alternative to traverse the matrix is to have multiple {\cuda} threads assigned to traversing different rows of the matrix, as each row can be computed concurrently with each other; see Figure~\ref{fig:m2lb}.
As with the diagonal-traversal but to a more moderate extent, the row-traversal can also be implemented so that it reuses a part of both the binomial and the power terms.

Let us consider for a moment a straightforward implementation to obtain the powers in Equation \eqref{eq:M2Lmat} shown in the code fragment in Figure~\ref{code:loop_divergence}. What characterizes this simple approach is that the number of iterations per thread will depend on $(k+n+1)$, which takes a different value for every thread. This straightforward implementation would have two problems: first, threads within a thread-block naturally stop at different points provoking the threads within a warp to diverge; second, loop unrolling would be unavailable as the compiler can only effectively unroll loops with known trip counts.

In our implementation, we separated the computation of powers in Equation \eqref{eq:M2Lmat} into two steps in order to minimize thread divergence, and to use compiler optimizations such as loop unrolling. In the first step of the calculation, the term $(x_{i} - x_{j})^{n+1}$ is computed by each thread; and in the second step, the next $k$  $(x_{i} - x_{j})$ factors, common to all threads, are multiplied. When computing the first step, we use a for-loop but with a fixed number of iterations ($P-1$, with $P$ the variable containing the truncation level of the {\ME}) and an \emph{if} conditional for computing only up to the power relevant to the thread; see the code fragment in Figure~\ref{code:loop_known_num_iter}. This approach minimizes thread divergence at the cost of a small overhead of evaluating a few extra empty iterations. However, the overhead is minimal when compared to the performance that would be lost due to thread divergence. In contrast to the first step, the second step consists of homogeneous computations for all threads, therefore it can be easily implemented using a for-loop with a known trip of $k$. The second step achieves high performance as it can be loop-unrolled and further optimized by the compiler. For a discussion on loop unrolling see \S\ref{unrolling}.

\subsubsection{Discussion of the {\ML} {\gpu} implementation and results}

\begin{table*}
\centering
\begin{tabular}{|c|c|c|c|c|c|c|c|c|}
\hline
\# of &  \# of  & {\ML} kernel & Reduction & Data HtD & Data DtH & OPS.   & B.          & Mill. trans. \\
 terms  &  trans. & [seconds]      & [seconds]  & [seconds] & [seconds]  & [Giga] & [GB/s] & per sec \\
\hline
\rowcolor[gray]{0.9} 8  &  2160  &  5.79e-05 &  3.48e-05 &  6.70e-05 &  3.10e-05 &  125.27 &  2.60 &  37.28  \\
\rowcolor[gray]{1.0} 8  &  9072  &  9.30e-05 &  5.51e-05 &  1.07e-04 &  4.01e-05 &  327.82 &  6.81 &  97.57  \\
\rowcolor[gray]{0.9} 8  &  36720  &  2.69e-04 &  1.41e-04 &  3.59e-04 &  8.99e-05 &  458.77 &  9.53 &  136.54  \\
\rowcolor[gray]{1.0} 8  &  147312  &  9.66e-04 &  4.91e-04 &  1.27e-03 &  2.80e-04 &  512.35 &  10.65 &  152.49  \\
\rowcolor[gray]{0.9} 8  &  589680  &  3.69e-03 &  1.91e-03 &  4.40e-03 &  8.29e-04 &  537.36 &  11.17 &  159.93  \\
\rowcolor[gray]{1.0} 8  &  2359152  &  1.44e-02 &  7.58e-03 &  1.65e-02 &  3.10e-03 &  548.72 &  11.40 &  163.31  \\
\hline
\rowcolor[gray]{0.9} 12  &  2160  &  7.41e-05 &  3.48e-05 &  6.91e-05 &  2.69e-05 &  185.97 &  2.93 &  29.13  \\
\rowcolor[gray]{1.0} 12  &  9072  &  1.60e-04 &  5.79e-05 &  1.17e-04 &  4.41e-05 &  362.02 &  5.71 &  56.71  \\
\rowcolor[gray]{0.9} 12  &  36720  &  5.11e-04 &  1.45e-04 &  4.09e-04 &  1.28e-04 &  458.60 &  7.24 &  71.84  \\
\rowcolor[gray]{1.0} 12  &  147312  &  1.90e-03 &  5.17e-04 &  1.37e-03 &  4.02e-04 &  493.93 &  7.79 &  77.37  \\
\rowcolor[gray]{0.9} 12  &  589680  &  7.44e-03 &  2.01e-03 &  4.87e-03 &  1.14e-03 &  506.32 &  7.99 &  79.31  \\
\rowcolor[gray]{1.0} 12  &  2359152  &  2.95e-02 &  8.00e-03 &  1.78e-02 &  4.51e-03 &  511.06 &  8.06 &  80.05  \\
\hline
\rowcolor[gray]{0.9} 16  &  2160  &  9.39e-05 &  3.48e-05 &  7.30e-05 &  2.91e-05 &  236.93 &  3.03 &  22.99  \\
\rowcolor[gray]{1.0} 16  &  9072  &  2.26e-04 &  5.70e-05 &  1.31e-04 &  5.20e-05 &  413.58 &  5.28 &  40.14  \\
\rowcolor[gray]{0.9} 16  &  36720  &  7.68e-04 &  1.52e-04 &  4.31e-04 &  1.62e-04 &  492.69 &  6.29 &  47.82  \\
\rowcolor[gray]{1.0} 16  &  147312  &  2.94e-03 &  5.36e-04 &  1.47e-03 &  5.15e-04 &  515.93 &  6.59 &  50.07  \\
\rowcolor[gray]{0.9} 16  &  589680  &  1.16e-02 &  2.08e-03 &  5.19e-03 &  1.55e-03 &  524.79 &  6.70 &  50.93  \\
\rowcolor[gray]{1.0} 16  &  2359152  &  4.61e-02 &  8.30e-03 &  1.89e-02 &  5.85e-03 &  526.90 &  6.73 &  51.14  \\
\hline
\end{tabular}
\caption{Results of the multipole-to-local computation on the {\NV} {\tesla} {\gpu}. Each row entry of the table presents
the results of a single test run. The description of the columns follows, from left to right:
number of terms computed, number of translations performed, {\gpu} execution time for {\ML} kernel,
{\gpu} execution time for reduction, time for data transfer from host to device (HtD), time for data transfer
from device to host (DtH), number of giga-operations per second (OPS.) for {\ML} kernel, effective bandwidth (B.)
utilization for {\ML} kernel, metric of {\ML} translations per second performed (in millions).}
\label{tab:m2ltesla}
\end{table*}

\begin{table*}
\centering
\begin{tabular}{|c|c|c|c|c|c|c|c|c|}
\hline
\# of &  \# of  & {\ML} kernel & Reduction & Data HtD & Data DtH & OPS.   & B.          & Mill. trans. \\
 terms  &  trans. & [seconds]      & [seconds]  & [seconds] & [seconds]  & [Giga] & [GB/s] & per sec \\
\hline
\rowcolor[gray]{0.9}8  &  2160  &  3.79e-05 &  2.10e-05 &  5.20e-05 &  2.19e-05 &  191.45 &  3.98 &  56.98  \\
\rowcolor[gray]{1.0}8  &  9072  &  7.61e-05 &  3.81e-05 &  9.92e-05 &  2.41e-05 &  400.79 &  8.33 &  119.28  \\
\rowcolor[gray]{0.9}8  &  36720  &  2.52e-04 &  1.25e-04 &  2.93e-04 &  4.89e-05 &  489.58 &  10.17 &  145.71  \\
\rowcolor[gray]{1.0}8  &  147312  &  9.41e-04 &  4.96e-04 &  8.46e-04 &  1.32e-04 &  526.11 &  10.93 &  156.58  \\
\rowcolor[gray]{0.9}8  &  589680  &  3.70e-03 &  2.02e-03 &  2.71e-03 &  5.79e-04 &  535.66 &  11.13 &  159.42  \\
\rowcolor[gray]{1.0}8  &  2359152  &  1.47e-02 &  8.19e-03 &  9.33e-03 &  2.55e-03 &  538.79 &  11.20 &  160.35  \\
\hline
\rowcolor[gray]{0.9}12  &  2160  &  5.10e-05 &  2.22e-05 &  4.72e-05 &  1.91e-05 &  270.27 &  4.26 &  42.34  \\
\rowcolor[gray]{1.0}12  &  9072  &  1.34e-04 &  4.01e-05 &  9.70e-05 &  2.79e-05 &  432.23 &  6.82 &  67.71  \\
\rowcolor[gray]{0.9}12  &  36720  &  4.87e-04 &  1.27e-04 &  2.65e-04 &  5.70e-05 &  481.27 &  7.59 &  75.39  \\
\rowcolor[gray]{1.0}12  &  147312  &  1.87e-03 &  5.03e-04 &  1.23e-03 &  2.98e-04 &  502.67 &  7.93 &  78.74  \\
\rowcolor[gray]{0.9}12  &  589680  &  7.42e-03 &  2.04e-03 &  2.86e-03 &  7.26e-04 &  507.56 &  8.01 &  79.50  \\
\rowcolor[gray]{1.0}12  &  2359152  &  2.96e-02 &  8.31e-03 &  9.34e-03 &  2.24e-03 &  509.26 &  8.03 &  79.77  \\
\hline
\rowcolor[gray]{0.9}16  &  2160  &  6.60e-05 &  2.22e-05 &  5.39e-05 &  2.00e-05 &  337.01 &  4.31 &  32.71  \\
\rowcolor[gray]{1.0}16  &  9072  &  1.88e-04 &  3.89e-05 &  1.17e-04 &  3.29e-05 &  497.56 &  6.36 &  48.29  \\
\rowcolor[gray]{0.9}16  &  36720  &  7.05e-04 &  1.31e-04 &  3.41e-04 &  7.70e-05 &  536.68 &  6.86 &  52.08  \\
\rowcolor[gray]{1.0}16  &  147312  &  2.73e-03 &  5.10e-04 &  1.27e-03 &  2.97e-04 &  556.61 &  7.11 &  54.02  \\
\rowcolor[gray]{0.9}16  &  589680  &  1.09e-02 &  2.08e-03 &  3.07e-03 &  9.81e-04 &  559.91 &  7.15 &  54.34  \\
\rowcolor[gray]{1.0}16  &  2359152  &  4.33e-02 &  8.43e-03 &  9.90e-03 &  2.94e-03 &  561.51 &  7.17 &  54.49  \\
\hline
\end{tabular}
\caption{Results of the multipole-to-local computation on the {\NV} {\fermi} {\gpu}. Each row entry of the table presents
the results of a single test run. The description of the columns follows, from left to right:
number of terms computed, number of translations performed, {\gpu} execution time for {\ML} kernel,
{\gpu} execution time for reduction, time for data transfer from host to device (HtD), time for data transfer
from device to host (DtH), number of giga-operations per second (OPS.) for {\ML} kernel, effective bandwidth (B.)
utilization for {\ML} kernel, metric of {\ML} translations per second performed (in millions).}
\label{tab:m2lfermi}
\end{table*}

One advantage of having more available threads per thread-block, is that all the memory transactions that take place to and from global memory can be organized so that threads in the same warp access sequential memory locations. This allows the {\gpu} to optimize the memory transfers by grouping together memory transactions to the same segment in global memory. In the row-traversal kernel, there are two cases when accessing global memory: the first case occurs when loading the multipole expansion coefficients from global memory to shared memory. The usage of a sequential memory layout of the \ME\ coefficients allows efficient memory transactions by warps of threads.  The second case occurs when the threads finish computing one local expansion coefficient, and here each thread directly saves the state into global memory. This global write is sequential between the threads, as at any given time a sequential number of coefficients are being computed by a sequential set of threads.

From a resource utilization point of view, each thread uses a small amount of resources: the shared memory usage is limited to storing the \ME\ information that is read at the start of the block execution, and all other memory is stored in the registers. Furthermore, as each thread only computes a single complex coefficient of a local expansion at a time, the overall register usage stays low.

\medskip

The crucial features of the row-traversal formulation of the kernel that resulted in remarkable performance gains are summarized as follows:

\begin{enumerate}
\item Increased the number of threads per block (in fact, we can have any number of threads).
\item Avoid thread branching for threads in the same warp.
\item Loop-unrolling (when done \emph{manually}, performance was even greater).
\item Reduced accesses to global memory.
\item When accessing global memory, we ensure it is \emph{coalesced}.
\end{enumerate}

The final step taken in the optimization of the {\gpu} formulation of the kernel was to overlap memory movements to and from global memory with computational work. The result is a high-performance algorithmic formulation tuned for the {\gpu} architecture, that attains \textbf{548} Giga operations per second on {\tesla} and up to \textbf{561} Giga operations per second on {\fermi}.

In  Table~\ref{tab:m2ltesla} and Table~\ref{tab:m2lfermi} we present performance results of several numerical experiments for the {\ML} kernel using {\tesla} and {\fermi} {\gpu}s, respectively. The numerical tests for the {\ML} implementation are controlled by two parameters: \emph{the number of terms of the multipole expansion}, and \emph{the number of expansions to be translated}. The intensity of the calculations is given by the number of terms of the {\ME}, and the size of the problem is given by the number of expansions to translate. Both tables report the time spent on the kernel execution and data transfer between host and {\gpu}, and performance metrics for the {\ML} kernel. Performance metrics reported correspond to the number of giga-operations per second, effective bandwidth utilization, and the number of {\ML} translations per second performed in millions.

Looking at the results presented in Table~\ref{tab:m2ltesla} and Table~\ref{tab:m2lfermi}, consider that the {\ML} is a compute-bound kernel. As such, it is limited by the maximum throughput of the {\gpu}. Therefore, for the {\gpu} to achieve its maximum throughput performance, it requires a test run large enough to completely utilize all of the streaming processors. This is reflected in the results shown, as the {\ML} kernel execution on the {\gpu} reports its best throughput on the test runs that have more {\ME} terms and larger problem sizes. We also report the number of translations per second as a ``real world'' performance metric that only depends on the test problem, \emph{i.e.}, it reflects wall-clock time to solution. Note that we obtained comparable single-precision performance between the {\tesla} and {\fermi} architectures. We attribute this to the fact that our application does not benefit from the new features introduced in {\fermi}, such as improved double precision performance, and L1 \& L2 cache.

%!TEX root = CruzLaytonBarba2010.tex
%%   fgt_gpu.tex

%%%  SUBSECTION
\subsection{Fast Gauss transform}

We will propose several improvements beyond simply rewriting {\cpu} code for {\cuda}, and later, in \S\ref{implementation} we will present actual code used for these accelerations. Firstly, instead of using the entire {\fgt} algorithm, we will focus on the evaluation of the Hermite series. According to many test runs with a full {\cpu} implementation of the {\fgt}, this part of the algorithms takes in the order of 90\% of the total run time,  so any optimizations here will have an appreciable effect on the overall speed of the algorithm. These optimizations will be placed in the context of the hardware metrics we introduced earlier in this section.

The Hermite series evaluation takes a set of series coefficients, $A_{\alpha}$, a target point (in $d$ dimensions), $y$, the center of the source cluster, $s_B$ and returns the influence of that source cluster at $y$. The equation form is given previously by \eqref{eqn:hermite}, but is presented again here for ease of reading:

\begin{equation}
G(y) = \sum_{\alpha \geq 0} A_{\alpha}h_{\alpha}\left ( \frac{y-s_B}{\sqrt{2}\sigma} \right ).
\end{equation}

\subsubsection{Reformulation of the {\fgt} algorithm for {\gpu}}

If one were to make no changes in the {\fgt}, then the {\gpu} implementation would follow the same logic as that used for a normal {\cpu} implementation. This means that for every source-target cluster interaction, a single {\cuda} kernel would be called. Each call uses the coefficients for this cluster along with its center point. Each thread would then take care of a single target point and compute the potential for that point from the Hermite coefficients. This approach, while naive, does have some advantages; it allows us to keep all variables needed by more than one thread in shared memory, while also ensuring concurrency and homogeneity are high. However, such an approach makes no deliberate attempt to leverage the advantages of {\cuda}. This is especially evident as the data sets being worked on are small\,---thus the overhead from memory accesses is going to be large compared to the run time, slowing down the implementation significantly.

Clearly, the approach described above will not give substantial acceleration on the hardware. Thus, we explore the algorithmic changes that will be necessary to improve performance. One first possibility is to work with multiple series expansions from multiple clusters at once. In this way, instead of sending the coefficients for a single source cluster, $A_\alpha$ and its associated cluster center, $s_B$, we amalgamate all necessary coefficients and cluster centers that must be evaluated at our target points into two large arrays, 

\begin{eqnarray}
	\label{eqn:amalgamated_coefficients}
	A_\alpha &=& \left [ A_\alpha^{(0)}, A_\alpha^{(1)}, \cdots, A_\alpha^{(j)} \right ]
\quad \text{and,} \\
		s_B & =&\left [ s_B^{(0)}, s_B^{(1)}, \cdots, s_B^{(j)} \right ]. 
\end{eqnarray}

\noindent In this way, we try to ``hide'' the time required to transfer all of the data to the device by increasing the computational intensity. We do this by performing a larger amount of work for each of the target points, which, once again, are assigned a single thread each, ensuring we keep the high levels of concurrency we would have enjoyed in a naive implementation.

This iteration falls short on a couple of new points: firstly, the number of values to be read into shared memory are too many. The size of the amalgamated coefficients, cluster centers and other necessary shared values such as $\alpha$, can exceed the limited amount of shared memory ($16$kB) available. Secondly, to maximize concurrency and occupancy, we choose to let each thread read only a single value and perform collaborative memory operations (as discussed in Section \S\ref{balancing}). The number of threads per block is hardware-limited to $512$, hence the number of shared values that can be read from global memory cannot be greater than this before we potentially compromise the homogeneity of the implementation. This indicates that we either need to store more values in global memory, take the performance hit and relinquish data-locality, or we must once again change our kernel.

%%% FINAL ITERATION %%%
\subsubsection{Final algorithm}\label{subsubsec:final_algo}
Taking into account the previously mentioned failings, our current implementation combines the best parts of both previous attempts to obtain very high performance. Instead of having one single kernel call with all the data needed, or multiple small calls as in a naive implementation, we split the evaluation into smaller calls. 
  Each call then  evaluates the influence of multiple source clusters, such that the amount of data needed is small enough to fit into shared memory, giving us spatial and temporal data-locality, and each thread need only read a single value into shared memory in order to ensure homogeneity. In this way, we ensure that all data needed across threads in a block (especially the series coefficients) is kept in shared memory, reducing the number of costly transfers between the device and host. This gives the {\gpu} enough data to take full advantage of its features, while ensuring that all memory accesses are as fast as possible.
  
 A further advantage of this approach is that spatial and temporal data-locality can be maximized. However, this refactoring is only one step of the optimization process. Within our evaluation of the Hermite series, we must perform other tasks, in particular evaluating Hermite polynomials for every appropriate source cluster at every target point.

% LOOP UNROLLING
\subsubsection{Hermite polynomial evaluation}

To evaluate polynomials in \cuda, we must first have some way of representing and working with them. Representing polynomials can be done with an ordered array of coefficients, but evaluation requires more work. There already exists an algorithmically optimal method for the evaluation in Horner's rule, and this method was implemented as a function in \cuda. Horner's rule allows us to evaluate a polynomial in $\mathcal{O}(n)$ time \cite{clrs2009}, where $n$ is the degree of the polynomial. It does this by writing the evaluation in the form:

\begin{equation}
	\label{eqn:horners}
	P(x) = \sum_{i=0}^{n-1}a_ix^i = a_0 + x(a_1 + x(a_2 + \cdots + x(a_{n-1})) \cdots).
\end{equation}

Our first version of this function used a loop over the polynomial coefficients, and was implemented as a device-only inlined function. This use of loops involves the evaluation of conditionals, and consequent potential branching of code. This branching causes significant performance impact in {\cuda}, and so this method needed to be re-implemented in some fashion to eliminate (if possible) all of the branching code.
Our solution to this problem was to manually unroll the evaluation loop, a process outlined in more detail in \S\ref{implementation}. This technique improved the computational intensity and thus throughput of the algorithm, and as an added incentive, reduced the number of registers that were being used, allowing us to optimize the occupancy more easily.

% SHARED MEMORY
\subsubsection{Shared memory use}\label{subsubsec:shared_mem}
The optimal use of shared memory is vitally important to overall performance. As outlined in \S\ref{gpuformulation}, the access time for shared memory is around $100\times$ faster compared to global memory. Thus, every variable that needs to be used by more than one thread, or even accessed more than once, should be copied into either shared memory (in the case of variables to be accessed by multiple threads) or local registers in order to obtain both spatial and temporal data locality. In the case of our implementation, the Hermite series coefficients, $A_\alpha^{(j)}$, the cluster centers, $s_B^{(j)}$, the multi-index values themselves, $\alpha$, all need to be copied into shared memory.

% COMPUTING ON THE FLY
\subsubsection{Computing on-the-fly}

A corollary of the access time for global memory is that one can have a situation where it is easier to calculate certain values on-the-fly, rather than read them from global memory. One such case is when calculating factorials. The standard method of calculating factorials by recursion is so computationally light that we can assign each factorial value to a single thread. This is preferential to the overhead of copying pre-computed values from global memory, and can be done while other threads are copying their values from global to shared memory. Additionally, all threads can calculate all values, but only save one. This allows us to unroll loops for greater performance and to maintain zero-branching.

% OCCUPANCY
\subsubsection{Occupancy}\label{subsubsec:occupancy}
Equally distributing work and memory accesses across threads is a necessary step for maximizing the occupancy of our implementation, as is reducing the number of registers used by each individual thread. These considerations allow us to make use of the highest possible number of threads at all times.

While the work performed by each thread is identical, ensuring homogeneity, memory transactions to global memory also need to be balanced. We achieve this balance by doing collective memory operations for retrieving data that is used by multiple threads. Each thread is then responsible for copying a single value from global memory into shared memory, and the index of the thread within its block dictates which value (polynomial coefficient, cluster center $x$ or $y$ value, etc.) the given thread will copy.   For a more complete discussion, see \S\ref{balancing}. 

% RESULTS
\subsubsection{Results}\label{subsubsec:fgt_results}
We present results of testing both a {\cuda} and a standard {\cpu} code implemented in C++. Attempts have been made to optimize this C++ code, including parallelization for shared memory systems using openMP, loop unrolling (using the \texttt{-funroll-loops} option from \texttt{gcc}) and evaluating hard-coded polynomial coefficients using Horner's rule. While this implementation is still not optimal, it offers a comparable code. This also explains the greater than $100\times$ speedups observed, as these should be questionable when comparing similarly optimal {\cpu} and {\gpu} codes.

% TIMING EXPLANATION
Timings for these tests were obtained by assuming that the kernels would be called within part of a greater algorithm. Thus, time taken to assign / populate target points and cluster co-efficients was not counted. In both the {\cpu} and {\gpu} cases these values would have been read from a file and computed by a separate method, respectively. 

\begin{table*}
\centering
\begin{tabular} { |c|c|c|c|c|c|c|}
\hline 
N & p & PPS & PPS & PPS & GOPS & PPS \\
 &  & (1 Core) & (2 Cores) & (4 Cores) & (GPU) & (GPU) \\ \hline
\rowcolor[gray]{0.9} 25600 & 5 & 1.68e+05 & 3.36e+05 & 6.59e+05 & 461.3 & 1.55e+07 \\ 
\rowcolor[gray]{1.0} 102400 & 5 & 1.68e+05 & 3.35e+05 & 6.60e+05 & 539.1 & 1.81e+07 \\ 
\rowcolor[gray]{0.9} 409600 & 5 & 1.69e+05 & 3.34e+05 & 6.68e+05 & 563.2 & 1.89e+07 \\ 
\rowcolor[gray]{1.0} 1638400 & 5 & 1.69e+05 & 3.37e+05 & 6.72e+05 & 570.7 & 1.91e+07 \\ 
\rowcolor[gray]{0.9} 6553600 & 5 & 1.68e+05 & 3.33e+05 & 6.72e+05 & 572.9 & 1.92e+07 \\ \hline
\rowcolor[gray]{1.0} 25600 & 9 & 3.99e+04 & 8.02e+04 & 1.59e+05 & 549.9 & 3.93e+06 \\ 
\rowcolor[gray]{0.9} 102400 & 9 & 4.03e+04 & 7.93e+04 & 1.58e+05 & 637.1 & 4.55e+06 \\ 
\rowcolor[gray]{1.0} 409600 & 9 & 4.03e+04 & 8.01e+04 & 1.59e+05 & 663.9 & 4.74e+06 \\ 
\rowcolor[gray]{0.9} 1638400 & 9 & 4.03e+04 & 8.03e+04 & 1.61e+05 & 671.6 & 4.80e+06 \\ 
\rowcolor[gray]{1.0} 6553600 & 9 & 4.02e+04 & 8.03e+04 & 1.60e+05 & 673.7 & 4.81e+06 \\ \hline
\rowcolor[gray]{0.9} 25600 & 12 & 2.05e+04 & 4.07e+04 & 8.08e+04 & 257.1 & 8.30e+05 \\ 
\rowcolor[gray]{1.0} 102400 & 12 & 2.06e+04 & 4.08e+04 & 8.16e+04 & 490.8 & 1.59e+06 \\ 
\rowcolor[gray]{0.9} 409600 & 12 & 2.04e+04 & 4.07e+04 & 8.18e+04 & 649.8 & 2.10e+06 \\ 
\rowcolor[gray]{1.0} 1638400 & 12 & 2.05e+04 & 4.09e+04 & 8.17e+04 & 688.8 & 2.23e+06 \\ 
\rowcolor[gray]{0.9} 6553600 & 12 & 2.05e+04 & 4.10e+04 & 8.15e+04 & 703.4 & 2.27e+06 \\ 
\hline
\end{tabular}
\caption{Results on both a multi-core processor and {\gpu} for the Hermite evaluation kernel. Presented are Particles per second (PPS) for all cases, and on the {\gpu} the total number of giga-operations (GOPS) obtained are also shown. Runs were made on a Penryn series Intel {\cpu} and {\NV} {\tesla} C1060 {\gpu}}
\label{tab:fgt_results_strange}
\end{table*}

\begin{table*}
\centering
\begin{tabular} { |c|c|c|c|c|c|c|}
\hline 
N & p & PPS & PPS & PPS & GOPS & PPS \\
 &  & (1 Core) & (2 Cores) & (4 Cores) & (GPU) & (GPU) \\ \hline
\rowcolor[gray]{0.9} 25600 & 5 & 2.31e+05 & 4.41e+05 & 7.11e+05 & 699.1 & 2.34e+07 \\ 
\rowcolor[gray]{1.0} 102400 & 5 & 2.50e+05 & 4.93e+05 & 9.39e+05 & 780.0 & 2.62e+07 \\ 
\rowcolor[gray]{0.9} 409600 & 5 & 2.54e+05 & 4.92e+05 & 9.31e+05 & 815.3 & 2.73e+07 \\ 
\rowcolor[gray]{1.0} 1638400 & 5 & 2.54e+05 & 4.98e+05 & 9.35e+05 & 828.7 & 2.78e+07 \\ 
\rowcolor[gray]{0.9} 6553600 & 5 & 2.58e+05 & 5.00e+05 & 9.76e+05 & 830.3 & 2.78e+07 \\ \hline
\rowcolor[gray]{1.0} 25600 & 9 & 7.12e+04 & 1.38e+05 & 2.64e+05 & 786.2 & 5.62e+06 \\ 
\rowcolor[gray]{0.9} 102400 & 9 & 7.24e+04 & 1.42e+05 & 2.65e+05 & 858.7 & 6.13e+06 \\ 
\rowcolor[gray]{1.0} 409600 & 9 & 7.19e+04 & 1.37e+05 & 2.78e+05 & 893.0 & 6.38e+06 \\ 
\rowcolor[gray]{0.9} 1638400 & 9 & 7.21e+04 & 1.39e+05 & 2.78e+05 & 905.9 & 6.47e+06 \\ 
\rowcolor[gray]{1.0} 6553600 & 9 & 7.29e+04 & 1.39e+05 & 2.80e+05 & 907.3 & 6.48e+06 \\ \hline
\rowcolor[gray]{0.9} 25600 & 12 & 3.33e+04 & 6.62e+04 & 1.20e+05 & 874.5 & 2.83e+06 \\ 
\rowcolor[gray]{1.0} 102400 & 12 & 3.33e+04 & 6.65e+04 & 1.27e+05 & 957.1 & 3.09e+06 \\ 
\rowcolor[gray]{0.9} 409600 & 12 & 3.32e+04 & 6.64e+04 & 1.29e+05 & 996.2 & 3.22e+06 \\ 
\rowcolor[gray]{1.0} 1638400 & 12 & 3.38e+04 & 6.52e+04 & 1.29e+05 & 1010.8 & 3.27e+06 \\ 
\rowcolor[gray]{0.9} 6553600 & 12 & 3.34e+04 & 6.64e+04 & 1.29e+05 & 1012.1 & 3.27e+06 \\ 
\hline
\end{tabular}
\caption{Results from the Hermite evaluation kernel on multi-core processors and {\gpu}. Particles per second (PPS) are presented for all cases, and the total number of giga-operations (GOPS) obtained on the {\gpu} are also given. Runs were made on a Nehalem series Intel {\cpu} and {\NV} {\fermi} C2050 {\gpu}}
\label{tab:fgt_results_yeager}
\end{table*}

\medskip

Tables \ref{tab:fgt_results_strange} and \ref{tab:fgt_results_yeager} show comparisons between {\cpu} and {\gpu} codes running on identical datasets with a number of evaluation points ranging up to over $6$ million. The truncation variable for the Hermite series is varied between  $p=3$ and $p=9$. For table \ref{tab:fgt_results_strange}, the {\cpu} implementation is run on between $1$ and $4$ cores of an Intel Penryn processor, and the {\gpu} code on a single \NV\ {\tesla} C1060 card. For table \ref{tab:fgt_results_yeager}, the processor is an Intel Nehalem and the {\gpu} is an {\NV} {\fermi} C2050. In both cases only the particles evaluated per second are compared (PPS in the tables), with the additional reporting of the number of giga-operations (GOPS) performed on the {\gpu} given as a measure of the total utilization of the card. Useful to note is that the {\cpu} code scales nearly linearly with the number of cores, showing it is at least fairly efficient, and that the greatest speedup from the {\gpu} with respect to $4$ cores on the {\gpu} is around $25$ times on the Nehalem / {\fermi} system and $30$ times for Penryn / {\tesla}. Given the good multi-processor scaling of the {\cpu} code, this represents a roughly $100-120$x speedup over a single {\cpu} core. The greater than $100\times$ speedup encountered is above what is commonly quoted as possible for a compute-bound application, but can be explained due to the possibility of many more optimizations (such as SSE operations) being implemented in the \cpu\ code. Particle evaluations per second (PPS) are also shown, to give a more ``physical'' representation of the speed of the code. While these results may seem high at first glance, they are within the theoretical peak of both cards (1033 Gops for the {\fermi} card), and we have verified by examining the intermediate .ptx files that fused multiply-add instructions have been used throughout the main inner loop that evaluates the Hermite polynomial. Given that our calculation is entirely throughput-bound, this ensures that we obtain very good performance. 

\medskip

The results presented show a significant disparity in the number of giga-operations performed, especially in the case of $p=12$, where the peak for the {\fermi} card is above $1000$ GOPS, while the {\tesla} card only manages $700$. The reason for this is likely the large number of integer operations required in the Hermite evaluation kernel in order to resolve memory locations. For instance, the full version of the code-unrolling snippet in Figure \ref{code:loops} requires only $2$ floating point operations ($+,  \times$), which can be done in a single fused multiply-add (FMA) instruction, while it requires an additional $2$ integer operations ($+,\times$) to obtain the correct coefficient memory location. In fact, for every floating point operation, on average, an integer operation must also be performed. Referring to Table 5-1 in {\NV}'s programming guide \cite{cuda-guide3.0}, we see that for the {\tesla} {\gpu}s (Compute Capability 1.3), 32-bit integer multiplies and multiply-adds both require multiple instructions, giving an integer throughput significantly lower than that for floating point operations. From the same table, we see that the {\fermi} card (Compute Capability 2.0) has the same throughput for these operations as for floating point operations. This significant increase in integer throughput removes a bottleneck in the code, and thus allows to improve our performance over the previous generation of cards.

%% SECTION 5
\section{GPU implementation discussion}\label{implementation}

%!TEX root = CruzLaytonBarba2010.tex

%%%% implementation details
In order to assist other practitioners in reaching these levels of performance, we now discuss the design strategies and in some cases demonstrate optimized fragments of code used within our kernels. These methods have been tuned for performance and are suitable for adaptation and use in any applicable {\cuda} implementation.

\subsection{Thread execution branching}

A code design consideration that greatly impacts performance is \emph{branching}\label{branching} of threads during execution due, for example, to data-dependent conditionals.  In the {\cuda} architecture, streaming multi-processors are able to manage a multitude of concurrent threads, but they do so in groups of $32$ parallel threads, called \emph{warps}.  All threads in a warp start together, but they may branch as they execute.  If they do, the warp actually executes in serial mode the different branch paths, while threads that do not take the particular path wait, effectively idling and wasting cycles.  Thus, all threads in the warp converge after the branching and then continue execution together.  For this reason, branching can be a serious hurdle for obtaining performance.

\subsection{Multithreading}

The {\cuda} architecture relies on no-cost multithreading to hide the high latency of memory accesses to global memory. A multiprocessor of a \textsc{gt}200 {\gpu} chip can have up to $1024$ active threads and $8$ active thread blocks at any given time. However, active threads on an {\sm} share its limited resources (registers and shared memory). In practice, for a given kernel implementation, the kernel's resource utilization defines the limit on the number of active threads per {\sm}. The ideal for a memory-bound application is to have as many active threads as possible, to hide the latency of the memory transactions. In order to characterize the efficiency of the {\sm} utilization, the concept of \emph{Occupancy}~\cite{cuda-guide} was introduced and defined as:
	
\begin{equation}
	\label{eqn:occupancy}
	\mathrm{Occupancy} = \frac{\mathrm{Number~of~active~threads}}{\mathrm{Maximum~threads~available}}
\end{equation}

The number of active threads will depend on a few parameters: on one hand, the kernel implementation will define the number of resources used per thread and per thread block, on the other hand, the resources per {\sm} are limited by the hardware. The only parameter that is defined by the user at run time is the number  of threads used per thread block. It is generally the case for memory-bound applications that by maximizing the occupancy of the multiprocessor we achieve better performance.
In our application, we maximized the {\gpu} occupancy in two ways: the most straightforward approach was to select a thread-block size that maximizes occupancy, one tool to do this is provided by \NV\ and is named ``occupancy calculator'' \footnote{\href{http://news.developer.nvidia.com/2007/03/cuda_occupancy_.html}{http://news.developer.nvidia.com/2007/03/cuda\_occupancy\_.html}}; the second approach was to hand-tune the implementation so that it reuses the maximum number of variables as possible. The last strategy has the most potential to improve occupancy by reducing the number of registers used, however it is a trial and error procedure, as the actual resource utilization ultimately depends on the compiler and sometimes less variables do not translate in less registers being used.

\subsection{Memory management}\label{subsec:mem_management}

Memory management needs to be explicitly provided by the programmer. This is a challenge, as the efficiency of the memory transactions directly depends on the algorithmic design and implementation. Issues that need to be considered are:
\begin{itemize}
\item [$\triangleright$] \emph{Small and fast shared memory}. The limited size of the shared memory, at 16 kB for the \gpu\ we used, clearly imposes a restriction on the amount of fast storage available. %number of threads in a thread block.  

\item [$\triangleright$] \emph{Large and slow global memory}. Global memory is characterized by the large memory space, with up to 4 GB of memory, but with very high latency of access from the \gpu\ chip (each transaction taking between 400 and 600 clock cycles).

\item [$\triangleright$] \emph{Shared memory conflicts}. Shared memory is physically divided into 16 blocks, and each block can do one memory transaction per clock (read, write, or broadcast). Therefore, if more than one thread of the same warp accesses data in the same memory block, the thread memory operations are queued and executed sequentially.

\item [$\triangleright$] \emph{Efficient global memory accesses}. Global memory is physically accessed as 32-, 64-, or 128-byte segments by the hardware. Therefore, every memory transaction that is issued will read or write to a whole segment, regardless of whether only one element of the segment is read or written into the segment. In this case, much of the memory bandwidth would be inefficiently used. Efficient memory transactions that are aligned and efficient per thread warp are commonly referred to as \emph{coalesced} memory transactions.

\item  [$\triangleright$] \emph{Memory camping}. Global memory is divided into physical blocks. The total bandwidth of global memory is calculated as the aggregated bandwidth of all the blocks. In order to effectively use the bandwidth, memory transactions need to be spread across all blocks. In our implementation, we addressed this by designing the kernels so that the thread blocks evenly access the global memory locations.

\end{itemize}

\subsection{Loop unrolling}\label{unrolling}

In several parts of our kernels, we require iterating over a loop to calculate values, for instance in evaluating multipole expansions in the {\fmm} and Hermite polynomials in the {\fgt}. Loop-unrolling is a well-known process whereby a loop over some pre-determined number of iterations, say $N$, is replaced with the loop's body code, repeated $N$ times. This technique trades a larger compiled binary size for increased performance.
This kind of optimization can be performed automatically by the {\cuda} compiler, but in our testing, we found that manually unrolling the code when possible resulted in both increased performance, and fewer registers being used, allowing us to more easily maximize the occupancy of our kernels.

The specific example we demonstrate is the evaluation of Hermite polynomials in the {\fgt}. This is done by implementing Horner's method, as previously described.  Consider the code example shown in Figure \ref{code:loops}. This code fragment shows the loop-based implementation, with \texttt{*H} being a pointer to an array of polynomial coefficients, \texttt{x} the desired evaluation point, and \texttt{poly\_len} denoting the degree of the polynomial to be evaluated. The code fragment in Figure \ref{code:unroll} demonstrates the same loop, but now unrolled for the instance where the polynomial degree is $5$.

%%%%% LOOP BASED CODE %%%%%
\begin{figure*}%[htbp]
\begin{center}
	\begin{lstlisting}
	__device__ float poly_eval(float *H, float x, int poly_len)
	{
	    int i;
	    float y = H[0];
	    
	    /* LOOP TO BE UNROLLED */
	    for (i=1; i < poly_len; i=i+1)
	    {
	        y = H[i] + x*y;
	    }
	
	    return y;
	}
	y = poly_eval(H,x,num_terms);
	\end{lstlisting}
\caption{Code snippet showing a loop-based implementation of the polynomial evaluation.}
\label{code:loops}
\end{center}
\end{figure*}

%%%%%% UNROLLED CODE %%%%%
\begin{figure*}%[htbp]
\begin{center}
	\begin{lstlisting}
	/* Manually unrolled code */
	y = H[0];
	y = H[1] + x*y;
	y = H[2] + x*y;
	y = H[3] + x*y;
	y = H[4] + x*y;
	\end{lstlisting}
\caption{Code snippet showing the unrolled loop.}
\label{code:unroll}
\end{center}
\end{figure*}

While this particular piece of code has been generated by hand,  it would be easy to have a separate piece of code to automatically generate at compile time the correct number of lines for a given number of terms. This would ensure the best possible performance, while removing the tedious and possibly error-prone manual adding of lines for more terms.

The optimization just described deals with speeding up computations, but we still need to get the data to compute efficiently. Thus, we now discuss balancing reads and writes from global memory.

\subsection{Balancing global reads/writes across threads}\label{balancing}

To maintain an equal amount of load across all working threads, it is important to ensure that no thread reads or writes disproportionately more to the global memory than any other. The main area where this problem appears is during the transfer of commonly-used variables from global to shared memory. A naive method may result in some threads reading multiple values into shared memory, while others read none, an obviously unsatisfactory situation. Thus, we propose the following system: the maximum number of values to be read into shared memory is the same as the number of threads within a block. The position of a thread within a block dictates which value it will read. This ensures that no thread reads more than one value. The example shown in Figure \ref{code:rw} is taken from our {\fgt} {\cuda} code.

\begin{figure*}%[htbp]
\begin{center}

	\begin{lstlisting}
	// shared memory
	__shared__ int shared_fact[NUM_TERMS];
	__shared__ int shared_alpha[LEN_ALPHA];
	__shared__ float shared_sb[2];
	
	// Read vars into shared memory
	// WARNING: Each block needs more threads than (LEN_ALPHA + NUM_TERMS + 2)

	if (threadIdx.x < NUM_TERMS) {
	    // generate factorial case
	    k = 0;
	    alpha1 = 1;
	        for (i=1; i < threadIdx.x + 1; i++) {
	            alpha1 *= i;
	        }
	        // store in shared memory
	        shared_fact[threadIdx.x + k] = alpha1;
	} else if (threadIdx.x < NUM_TERMS + LEN_ALPHA) {
	    // read alpha into shared
	    k = -NUM_TERMS;
	    shared_alpha[threadIdx.x + k] = alpha[threadIdx.x + k];
	} else if (threadIdx.x < NUM_TERMS + LEN_ALPHA + 2) {
	    // read sb into shared memory
	    k = -NUM_TERMS - LEN_ALPHA;
	    shared_sb[threadIdx.x + k] = sb[threadIdx.x + k];
	} else {
	    // default case - no read
	    k = 0;
	}
	__syncthreads();
	\end{lstlisting}

\caption{Code snippet showing the balanced reads/writes.}
\label{code:rw}
\end{center}
\end{figure*}

The code snippet  divides our threads so that as many values are read simultaneously as possible, with the type of variable to be read determined by the thread's index within its block. The first group of threads will calculate factorial values; the second will read polynomial coefficients and the third group will read the expansion center. In the second part of the code snippet, lines $29$ to $49$, the code segment diverges the reads across all threads and contiguous reads are achieved by having adjacent threads reading adjacent values from memory. It is important to note here that in earlier versions of the {\NV} {\cuda} compiler, ``if'' statements forced thread branching on a whole threadblock, while a ``switch'' statement only branched within a thread warp. Thus, the optimal version of the code in Figure \ref{code:rw} for compilers previous to version $3.x$ involved splitting the code into two separate steps: an ``if'' statement to decide which threads read which values, and then a separate ``switch'' step to actually perform the reads. However, with compiler versions 3.0 and above, this is no longer appropriate, and so the code in Figure \ref{code:rw} is optimal.

This implementation has significant advantages: first, memory transactions can be organized such that adjacent threads within the same warp can read/write adjacent values from/to global memory, therefore ensuring coalesced memory transactions. Secondly, while this example only handles reading as many values as we have threads, it is trivially extendable to handle as many reads as a user desires. Finally, for some memory-intensive kernels, it is possible to define the size of a thread block to obtain the best kernel performance for memory transactions, even if this implies that some threads could remain idle during the compute-intensive part of the kernel.

\subsection{Other design strategies}\label{onfly}

\begin{itemize}

\item [$\triangleright$] \emph{Keep the kernel complexity low}. It is better to have a highly specialized kernel than a general purpose implementation. We consider three reasons for this: first, highly specialized kernels keep the resource utilization low, which in turn allows having more active threads per multiprocessor; second, high specialization allows for the use of specific optimizations; third, general cases normally result in branching conditionals, and this reduces the kernel performance.

\item [$\triangleright$] \emph{Use many threads per thread block}. Many threads can cooperate during memory transactions, resulting in more efficient use of the memory bandwidth. This strategy is more important in the negative, \emph{i.e.}, for kernels that use \emph{too few} threads it will be difficult to perform coalesced memory transactions.

\item [$\triangleright$] \emph{Interleave computations and memory transactions}. In a kernel, it is a good idea to interleave memory transfers with computational work within the same kernel, in contrast with using a three-stage kernel that consists of one stage to load data, a second stage to perform computations, and a third stage for saving data. We observed that in our cases it was better to have smaller stages that interleave memory transfers (load/write) with work. In this way, the multiprocessor-limited resources can be used more efficiently and it is easier to hide latency by multithreading, as it is more difficult to overlap threads performing big stages than a series of small efficient stages.

\item [$\triangleright$]  \emph{On-the-fly calculations}. As stated several times before, there is a significant overhead related to reading and writing from global memory. This overhead can be so significant that for some purposes it may be more efficient to re-calculate values than to store and retrieve them. A good example of this is found in a factorial cache. In a standard {\cpu} code, it is more efficient to pre-compute the values of all the factorials you may need to use at the beginning of your code, then simply re-use these values when needed. In a {\gpu} implementation however, these values would need to be read into shared memory for each block. Instead of doing this, we consider designating a set of threads in each block to calculate these values instead of reading from memory. 

\end{itemize}

\section{Conclusions} 
\label{conclusions}

This work demonstrates that it is possible to attain close to the practical peak of modern {\gpu} architectures, not just with embarrassingly parallel problems, but with real-world, elaborate algorithms.  The process, however, is not straight-forward.  Intimate knowledge of the architectural features is required to rethink the core methods such that performance can be maximized.  This familiarity with the hardware architecture, in fact, reveals that some algorithms will be better suited to perform well on the {\gpu}. Some others will suffer some unavoidable bottlenecks due to being bound by memory bandwidth, rather than computation. When this happens, only moderate speedups will be possible on the new hardware. When a computationally intensive algorithm is properly reformulated for the {\gpu}, however, two-order of magnitude speedups can be obtained.  This holds fantastic promise for extending the capability of computing to solve some of the most challenging scientific problems.

The results obtained here, sustaining over $500$ Gigaop/s on one {\tesla} C1060 card for fast summation algorithms, can have significant impact for applications using these algorithms.  In particular, we are currently working to develop boundary element methods which are accelerated with the fast multipole method. Considering the potential speedup in applications, we anticipate unprecedented capability for acoustics, electromagnetics, bioelectrostatics, and other applications of the \textsc{bem} and {\fmm}.

Our next challenge is to utilize the {\gpu} kernels developed for this work in a truly heterogeneous environment.  We will couple these kernels to software libraries for distributed systems, and investigate scalability with multiple {\cpu} cores and multiple {\gpu}s.  Such work will offer application scientists tools for truly groundbreaking discovery through advanced computing.

\bigskip

Codes and test scripts used to obtain the results presented in this paper are available from \href{http://code.google.com/p/cufast/}{http://code.google.com/p/cufast/}.

\section*{Acknowledgments}
FAC funded by Airbus and BAE Systems under contract ACAD 01478. SKL funded by a Boston University College of Engineering Dean's Fellowship.
LAB acknowledges partial support from EPSRC under grant contract EP/E033083/1, and from Boston University College of Engineering.
Thanks to Mike Clark for a careful reading of an early version of the manuscript, and thanks to Nvidia for an equipment donation to the BU Center for Computational Science that aided in this work.

%%  BIBLIOGRAPHY

\bibliographystyle{elsarticle-num}

%\bibliography{/Users/felipe/Documents/repositories/bibtex/FastMethods,/Users/felipe/Documents/repositories/bibtex/scicomp,/Users/felipe/Documents/repositories/bibtex/vortexflow,/Users/felipe/Documents/repositories/bibtex/vortexmeth,/Users/felipe/Documents/repositories/bibtex/meshfree,/Users/felipe/Documents/repositories/bibtex/biomodeling}
\bibliography{FastMethods,scicomp,vortexmeth}

%\begin{thebibliography}{00}

% \bibitem{label}
% Text of bibliographic item

% notes:
% \bibitem{label} \note

% subbibitems:
% \begin{subbibitems}{label}
% \bibitem{label1}
% \bibitem{label2}
% If there is a note, it should come last:
% \bibitem{label3} \note
% \end{subbibitems}

%\bibitem{}

%\end{thebibliography}

\end{document}